\documentclass[12pt]{article}
\usepackage{amssymb, latexsym, amsmath, hyperref}

\begin{document}

\title{\textbf{Pseudoduality In Supersymmetric Sigma
Models}}

\author{\textbf{Mustafa Sarisaman}\footnote{msarisaman@physics.miami.edu}}

\date{}

\maketitle
\par
\begin{center}
\textit{Department of Physics\\ University of Miami\\ P.O. Box 248046\\
 Coral Gables, FL 33124 USA}
\end{center}

\

\begin{center}
Monday, April 27, 2009
\end{center}

\

\begin{abstract}
We study the pseudoduality transformation in supersymmetric sigma
models. We generalize the classical construction of pseudoduality
transformation to supersymmetric case. We perform this both by
component expansion method on manifold $\mathbb{M}$ and by
orthonormal coframe method on manifold $SO(\mathbb{M})$. The
component expansion method yields the result that pseudoduality
tranformation is not invertible at all points and occurs from all
points on one manifold to only one point where riemann normal
coordinates are valid on the second manifold.  Torsion of the sigma
model on $\mathbb{M}$ must vanish while it is nonvanishing on
$\tilde{\mathbb{M}}$, and curvatures of the manifolds must be
constant and the same. In case of super WZW sigma models
pseudoduality equations result in three different pseudoduality
conditions; flat space, chiral and antichiral pseudoduality.
\end{abstract}

\vfill

\section{Introduction}\label{sec:int}

This model consists of both bosons and fermions, and they are
transformed into each other by supersymmetry transformation. It
improves the short distance behaviour of quantum theories and gives
a beautiful solution to the hierarchy problem. Supersymmetric sigma
models have a rich geometrical structure. It has been shown
\cite{ketov1} that target space of $N = 1$ sigma models is a
(pseudo-)Riemannian manifold, $N = 2$ is the K\"{a}hler manifold and
$N = 4$ is the hyper-K\"{a}hler manifold. Sigma models based on
manifolds with torsion \cite{hullwitten} have chiral supersymmetry
in which the number of left handed supersymmetries differs from the
number of right handed supersymmetries.

There is an interesting duality transformation proposed by authors
\cite{curtright1, ivanov, alvarez1, alvarez2}, which is called as
"pseudoduality". By contrast with usual duality transformations this
"on shell duality" transformation is not canonical, and maps
solutions of the equations of motion of the "pseudodual" models. We
will use the term pseudodual when there is a pseudoduality
transformation between different models. It is pointed out that this
transformation preserves the stress energy tensor \cite{alvarez2}.

In \cite{alvarez2}, pseudoduality in classical sigma models was
extensively discussed, and in this paper we are going to analyze
pseudoduality transformation of supersymmetric extension of
classical sigma models. We will focus on (1,0) and (1, 1) real
supersymmetric sigma models in two dimensions, and find the required
conditions which supersymmetry constrains the target space and
following results for pseudoduality. We will refer to references
\cite{ketov1, wess1, west1, gates1, freund1} about supersymmetry and
superspace constructions.

We use the superspace coordinates $(\sigma^{\pm}, \theta^{\pm})$,
where the bosonic coordinates $\sigma^{\pm} = \tau\pm\sigma$ are the
usual lightcone coordinates in two-dimensional Minkowski space, and
the fermionic coordinates  $\theta^{\pm}$ are the Grassmann numbers.
The supercovariant derivatives are
\begin{equation} \label{equation4.1}
D_{\pm} = \partial_{\theta^{\pm}} + i\theta^{\pm}\partial_{\pm}
\end{equation}
and the supercharges generating supersymmetry are
\begin{equation} \label{equation4.2}
Q_{\pm} = \partial_{\theta^{\pm}} - i\theta^{\pm}\partial_{\pm}
\end{equation}
and it follows that
\begin{equation} \label{equation4.3}
Q_{\pm}^{2} = - i\partial_{\pm} \ \ \ \ \ \ \ D_{\pm}^{2} =
 i\partial_{\pm}.
\end{equation}
and all other anticommutations vanish. The scalar superfields in
components have the form
\begin{equation} \label{equation4.4}
X (\sigma, \theta) = x (\sigma) + \theta^{+}\psi_{+}(\sigma) +
\theta^{-}\psi_{-}(\sigma) + \theta^{+}\theta^{-}F (\sigma)
\end{equation}
where $x : \Sigma \rightarrow M$, $\psi_{\pm}$ are the two
dimensional Majorana spinor fields, and $F$ is the auxiliary real
scalar field.

\section{Pseudoduality in Heterotic Sigma Models}\label{sec4:PHSM}

This model \cite{ketov1, hullwitten, zandron, delduc,
michalogiorgakis, hull} is enlarging the spacetime $\Sigma$ in the
classical case to the superspace $\Xi^{1, 0}$ by adding a Grassmann
degree of freedom. Hence the sigma model is the map consisting of a
scalar $x$ and a fermion $\psi_{+}$. This case has one left-handed
supercharge $Q_{+}$, and does not contain any right-handed
supercharge $Q_{-}$. The supersymmetry algebra will be
\begin{equation}
\{Q_{+}, Q_{+}\} = 2 i P_{+} \notag
\end{equation}
where $\{ , \}$denotes anticommutation, and $P_{+} = - \partial_{+}$
as can be checked from (\ref{equation4.3}). The supersymmetry
transformations generated by $Q_{+}$ will be
\begin{align}
\delta_{\epsilon} x (\sigma) &= \epsilon_{-} \psi_{+} (\sigma)
\notag\\
\delta_{\epsilon} \psi_{+} (\sigma) &= i \epsilon_{-} \partial_{+} x
(\sigma) \notag
\end{align}
Hence the fermion $\psi_{+}$ can be thought of as the superpartner
of the boson $x$. In what follows we will examine pseudoduality
transformations  between supermanifolds $\mathbb{M}$
\footnote{$\mathbb{M}$ is the target space in which supersymmetric
sigma models is defined.} and $\tilde{\mathbb{M}}$ using components
first, and then probe how it behaves when lifted to orthonormal
coframe bundles $SO(\mathbb{M})$ \footnote{$SO(\mathbb{M}) =
\mathbb{M} \times SO(n)$.} and $SO(\tilde{\mathbb{M}})$. We
emphasize that pseudoduality is defined between superspaces $z$
which are the pullbacks of the manifols $\mathbb{M}$ and
$\tilde{\mathbb{M}}$ in case of components, and $SO(\mathbb{M})$ and
$SO(\tilde{\mathbb{M}})$ in case of orthonormal coframe method. This
is implicitly intended in our calculations.

\subsection{Components}

In this case the superfield $X$ has the form
\begin{equation} \label{equation4.5}
X = x (\sigma) + \theta^{+}\psi_{+}(\sigma)
\end{equation}
where $X : \Xi^{1, 0} \rightarrow \mathbb{M}$, and $\Xi^{1, 0} =
(\sigma^{+}, \sigma^{-}, \theta^{+})$. The real grassmann coordinate
$\theta^{+}$ is anticommuting and $(\theta^{+})^{2} = 0$. We will
assume that target space has torsion H, which is introduced into the
action by a Wess-Zumino term. Reparametrization invariant action
defined on a Riemannian manifold $\mathbb{M}$ with metric $G_{ij}$,
standard connection $\mathbf{\Gamma}^{i}_{jk}$ and antisymmetric
two-form $B_{ij}$ can be written as
\begin{equation} \label{equation4.6}
S = \int d^{2}\sigma d \theta (G_{ij} + B_{ij})
D_{+}X^{i}\partial_{-}X^{j}
\end{equation}

We may write similar expressions for manifold $\tilde{\mathbb{M}}$
using expressions with tilde. Since we want to write down
pseudoduality transformations between two manifolds, we need to find
out the equations of motions from action (\ref{equation4.6}). If we
write this action in terms of bosonic coordinates of superspace
only, we obtain our original classical action plus fermionic terms.
After expanding $G_{ij}$ and $B_{ij}$ in the first order terms and
integrating this action under $d\theta$ gives the following
\begin{equation} \label{equation4.7}
S = \int d^{2}\sigma [i(g_{ij} +
b_{ij})\partial_{+}x^{i}\partial_{-}x^{j} -
g_{ij}\psi_{+}^{i}\nabla_{-}^{(-)}\psi_{+}^{j}]
\end{equation}
where $\nabla_{-}^{(-)}\psi_{+}^{j} = \nabla_{-}\psi_{+}^{j} -
H^{j}_{kl}\psi_{+}^{k}\partial_{-}x^{l}$ and $\nabla_{-}\psi_{+}^{j}
= \partial_{-}\psi_{+}^{j} +
\Gamma^{j}_{kl}\psi_{+}^{k}\partial_{-}x^{l}$, and $H_{ijk} =
\frac{1}{2}(\partial_{i}b_{jk} + \partial_{j}b_{ki} +
\partial_{k}b_{ij})$. Equations of motion following from the action
(\ref{equation4.7}) are
\begin{align} \label{equation4.8}
\nabla_{-}^{(-)}\psi_{+}^{i} &= 0 \\
\square x^{k} &=
i\hat{R}^{k}_{lij}\psi_{+}^{i}\psi_{+}^{j}\partial_{-}x^{l}
\label{equation4.9}
\end{align}
where $\square x^{k} = \nabla_{+}^{(+)}\partial_{-}x^{k} +
\nabla_{-}^{(-)}\partial_{+}x^{k}$, and the generalized curvature is
defined as
\begin{equation}\label{equation4.10}
\hat{R}_{ijkl} = R_{ijkl} - D_{k}H_{ijl} + D_{l}H_{ijk} +
H_{ikn}H^{n}_{lj} - H_{jkn}H^{n}_{li}
\end{equation}

We can write the Pseudoduality transformations as follows
\begin{align} \label{equation4.11}
D_{+}\tilde{X}^{i} &= + \mathcal{T}_{j}^{i} D_{+}X^{i} \\
\partial_{-}\tilde{X}^{i} &= - \mathcal{T}_{j}^{i} \partial_{-}X^{j}
\label{equation4.12}
\end{align}
where $\mathcal{T}$ is the transformation matrix, and is a function
of superfield $X$. Since superfield depends on $\sigma$ and
$\theta^{+}$, we may say that $\mathcal{T}$ is a function of
$\sigma$ and $\theta^{+}$. We let $\mathcal{T}(\sigma, \theta) =
T(\sigma) + \theta^{+}N(\sigma)$. Splitting pseudoduality equations
into the fermionic and bosonic parts leads to the following set of
equations
\begin{align} \label{equation4.13}
\tilde{\psi}_{+}^{i}(\sigma) &= + T_{j}^{i}(\sigma) \psi_{+}^{j}(\sigma) \\
\partial_{-}\tilde{\psi}_{+}^{i} (\sigma) &= - T_{j}^{i} (\sigma)
\partial_{-}\psi_{+}^{j}(\sigma) - N_{j}^{i}(\sigma)
\partial_{-}x^{j}(\sigma) \label{equation4.14}\\
\partial_{+}\tilde{x}^{i}(\sigma) &= + T_{j}^{i}(\sigma)
\partial_{+}x^{j}(\sigma) - iN_{j}^{i}(\sigma)\psi_{+}^{j}(\sigma)
\label{equation4.15}
\\
\partial_{-}\tilde{x}^{i} (\sigma) &= - T_{j}^{i} (\sigma)
\partial_{-}x^{j} (\sigma) \label{equation4.16}
\end{align}

We see that the component $T$ is responsible for the classical
transformation which does not change the type of field, while $N$
contributes to the fermionic degree of transformation which
transforms bosonic fields to fermionic ones, and vice versa. Before
finding pseudodual expressions it is worth to obtain constraint
relations. We take $\partial_{-}$ of (\ref{equation4.13}) and set
equal to (\ref{equation4.14}), and then use the equation of motion
(\ref{equation4.8}) to obtain
\begin{equation} \label{equation4.17}
N_{k}^{i} = - [M_{lk}^{i} + 2T_{j}^{i}(H_{lk}^{j} -
\Gamma_{lk}^{j})]\psi_{+}^{l}
\end{equation}
where we define $\partial_{k}T_{l}^{i} = M_{lk}^{i}$. Now taking
$\partial_{+}$ of (\ref{equation4.16}) and setting equal to
$\partial_{-}$ of (\ref{equation4.15}) followed by using equations
of motion (\ref{equation4.8}) and (\ref{equation4.9}) yields
\begin{align} \label{equation4.18}
&[2T_{k}^{i}(H_{mn}^{k} - \Gamma_{mn}^{k}) + 2M_{(mn)}^{i}]
\partial_{+}x^{m}\partial_{-}x^{n} + iT_{k}^{i}\hat{R}^{k}_{mij}
\psi_{+}^{i} \psi_{+}^{j}
\partial_{-}x^{m} \notag\\ &= iN_{k}^{i}(H_{mn}^{k} - \Gamma_{mn}^{k})
\psi_{+}^{m} \partial_{-}x^{n} +
i(\partial_{-}N_{k}^{i})\psi_{+}^{k}
\end{align}
where $M_{(mn)}^{i}$ represents the symmetric part of $M_{mn}^{i}$.
Real part of this equation gives
\begin{equation}
T_{k}^{i}(H_{mn}^{k} - \Gamma_{mn}^{k}) + 2M_{(mn)}^{i} = 0
\label{equation4.19}
\end{equation}
which implies that
\begin{align} \label{equation4.20}
H_{mn}^{k} &= 0, \\
M_{(mn)}^{i} &= T_{k}^{i}\Gamma_{mn}^{k} \label{equation4.21}
\end{align}
Substituting these results into (\ref{equation4.17}) leads to
\begin{equation} \label{equation4.22}
N_{k}^{i} = M_{km}^{i} \psi_{+}^{m}
\end{equation}

Complex part of (\ref{equation4.18}) together with
(\ref{equation4.20}), (\ref{equation4.21}) and (\ref{equation4.22})
gives the following equation
\begin{equation} \label{equation4.23}
\partial_{n}M_{[mj]}^{i} = T_{k}^{i}R_{njm}^{k} + 2M_{[kj]}^{i}
\Gamma_{mn}^{k}
\end{equation}
where $M_{[mj]}^{i}$ denotes the antisymmetric part of $M_{mj}^{i}$.
Solution of this equation gives the result for $T$.

\subsubsection{Riemann Normal Coordinates} \label{sec4:RNC}

Before we attempt to find the general (global) solution for the
equation (\ref{equation4.23}), it is interesting to find the special
solution where Riemann Normal coordinates \cite{carroll1, agapitos,
greenschwartzwitten} are used in both models. In these coordinates
solution is expanded around a point (call this point as $p$ on $M$,
and $\tilde{p}$ on $\tilde{M}$) which Christoffel's symbols vanish.
Curvature tensor $R$ is the curvature of the point $p$, and
constant. (\ref{equation4.21}) implies that $M_{jm}^{i} = -
M_{mj}^{i}$, and hence, equation (\ref{equation4.23}) is reduced to
\begin{equation}
\partial_{n}M_{mj}^{i} = T_{k}^{i}R_{njm}^{k} \notag
\end{equation}
After integration we get
\begin{equation} \notag
M_{mj}^{i} = M_{mj}^{i} (0) + \int T_{k}^{i}R^{k}_{njm} dx^{n}
\end{equation}
and since $T_{m}^{i} = T_{m}^{i} (0) + \int M_{mj}^{i} dx^{j}$, we
finally obtain
\begin{equation}
T_{m}^{i} = T_{m}^{i} (0) + M_{mj}^{i} (0) x^{j} + T_{k}^{i} (0)
R^{k}_{njm} \int x^{n} dx^{j} + M_{kl}^{i} (0) R^{k}_{njm} \int
dx^{j} \int x^{l} dx^{n} + H.O. \notag
\end{equation}
and
\begin{equation}
M_{mj}^{i} = M_{mj}^{i} (0) + T_{k}^{i} (0) R^{k}_{njm} x^{n} +
M_{kl}^{i} (0) R^{k}_{njm} \int x^{l} dx^{n} + H.O. \notag
\end{equation}
and also using (\ref{equation4.22}) we find
\begin{equation}
N_{k}^{i} = M_{km}^{i} (0) \psi_{+}^{m} + T_{j}^{i} (0) R^{j}_{nmk}
\psi_{+}^{m} x^{n} + M_{jl}^{i} (0) R^{j}_{nmk} \psi_{+}^{m} \int
x^{l} dx^{n} + H.O. \notag
\end{equation}
We choose the initial condition $T_{m}^{i} (0) = \delta_{m}^{i}$.
Hence Pseudoduality relations (\ref{equation4.13}) -
(\ref{equation4.16}) up to the second order in $x$ can be written as
\begin{align} \label{equation4.24}
\tilde{\psi}_{+}^{i} &= \psi_{+}^{i} + M_{jk}^{i} (0) \psi_{+}^{j}
x^{k} + R^{i}_{nkj} \psi_{+}^{j} \int x^{n} dx^{k} + H.O. \\
\partial_{-} \tilde{\psi}_{+}^{i} &= - M_{jm}^{i} (0) \psi_{+}^{m}
\partial_{-}x^{j} - R^{i}_{nmj} \psi_{+}^{m} x^{n} \partial_{-}x^{j}
+ H.O. \label{equation4.25} \\
\partial_{+}\tilde{x}^{i} &= \partial_{+}x^{i} + M_{jk}^{i} (0) x^{k}
\partial_{+}x^{j} - iM_{jm}^{i} (0) \psi_{+}^{m} \psi_{+}^{j}
-iR^{i}_{nmj} \psi_{+}^{m} \psi_{+}^{j} x^{n} \notag\\ &+
R^{i}_{nkj} \partial_{+}x^{j} \int x^{n} dx^{k} -i M_{kl}^{i} (0)
R^{k}_{nmj} \psi_{+}^{m} \psi_{+}^{j} \int x^{l} dx^{n} + H.O.
\label{equation4.26}\\
\partial_{-}\tilde{x}^{i} &= - \partial_{-} x^{i} - M_{jk}^{i} (0)
x^{k} \partial_{-}x^{j} - R^{i}_{nlj} \partial_{-}x^{j} \int x^{n}
dx^{l} + H.O. \label{equation4.27}
\end{align}

Using the equation of motion (\ref{equation4.8}) for tilde, i.e.
$\partial_{-}\tilde{\psi}_{+}^{i} = \tilde{H}_{jk}^{i}
\tilde{\psi}_{+}^{j} \partial_{-}\tilde{x}^{k}$, and combining with
(\ref{equation4.24}) and (\ref{equation4.27}) we find
\begin{equation} \label{equation4.28}
\partial_{-} \tilde{\psi}_{+}^{i} = - \tilde{H}_{mj}^{i}
\psi_{+}^{m} \partial_{-}x^{j} - \tilde{H}_{mk}^{i} M_{jn}^{k} (0)
\psi_{+}^{m} x^{n} \partial_{-}x^{j} - \tilde{H}_{kj}^{i} M_{mn}^{k}
(0) \psi_{+}^{m} x^{n} \partial_{-}x^{j} + H.O.
\end{equation}
A comparison of equation (\ref{equation4.25}) with equation
(\ref{equation4.28}) gives
\begin{align} \label{equation4.29}
\tilde{H}_{mj}^{i} &= M_{jm}^{i} (0) \\
R^{i}_{nmj} &= M_{km}^{i} (0) M_{jn}^{k} (0) + M_{jk}^{i} (0)
M_{mn}^{k} (0) \label{equation4.30}
\end{align}
Now we see that equation (\ref{equation4.9}) with tilde is written
as $\partial^{2}_{+-} \tilde{x}^{i} = \tilde{H}_{jk}^{i}
\partial_{+}\tilde{x}^{j} \partial_{-}\tilde{x}^{k} +
\frac{i}{2}\hat{\tilde{R}}^{i}_{jkl} \tilde{\psi}_{+}^{k}
\tilde{\psi}_{+}^{l} \partial_{-}\tilde{x}^{j}$. Inserting
(\ref{equation4.24}), (\ref{equation4.26}) and (\ref{equation4.27})
into this equation gives
\begin{equation} \label{equation4.31}
\partial^{2}_{+-}\tilde{x}^{i} = - \tilde{H}_{jk}^{i}
\partial_{+}x^{j} \partial_{-}x^{k} + i\tilde{H}_{jk}^{i} M_{mn}^{j}
(0) \psi_{+}^{n} \psi_{+}^{m} \partial_{-}x^{k} - \frac{i}{2}
\hat{\tilde{R}}^{i}_{jkl} \psi_{+}^{k} \psi_{+}^{l}
\partial_{-}x^{j} + H.O.
\end{equation}
Likewise we can write a relation for
$\partial^{2}_{+-}\tilde{x}^{i}$ using (\ref{equation4.26}) or
(\ref{equation4.27}) as
\begin{equation} \label{equation4.32}
\partial^{2}_{+-}\tilde{x}^{i} = - M_{kj}^{i} (0) \partial_{+} x^{j}
\partial_{-} x^{k} - \frac{i}{2} R^{i}_{jkl} \psi_{+}^{k}
\psi_{+}^{l} \partial_{-}x^{j} + H.O.
\end{equation}
A simple comparison of (\ref{equation4.31}) with
(\ref{equation4.32}) gives the following
\begin{align} \label{equation4.33}
\tilde{H}_{jk}^{i} &= M_{kj}^{i} (0) \\
-R^{i}_{jkl} &= - \hat{\tilde{R}}^{i}_{jkl} + 2\tilde{H}_{nj}^{i}
\tilde{H}_{kl}^{n} \label{equation4.34}
\end{align}

We notice that (\ref{equation4.29}) is the same as
(\ref{equation4.33}), and $- \hat{\tilde{R}}^{i}_{jkl} +
2\tilde{H}_{nj}^{i} \tilde{H}_{kl}^{n} = - \tilde{R}^{i}_{jkl}$.
Therefore we obtain $R^{i}_{jkl} = \tilde{R}^{i}_{jkl}$. We see that
curvatures of the points $p$ and $\tilde{p}$ are constant and same.
This implies that pseudoduality between two models based on Riemann
normal coordinates must have same curvatures. We see from
(\ref{equation4.29}) and (\ref{equation4.30}) that this
transformation works in one way, and is not invertible in this
special solution.

\subsubsection{General Solution} \label{sec4:GS}

Now we find the global solution to equation (\ref{equation4.23}). We
know that we can write $M_{kj}^{i}$ as the sum of symmetric and
antisymmetric parts as follows
\begin{equation}
M_{kj}^{i} = \frac{1}{2}(M_{kj}^{i} - M_{jk}^{i}) +
\frac{1}{2}(M_{kj}^{i} + M_{jk}^{i}) \notag
\end{equation}
Inserting antisymmetric part of this matrix into
(\ref{equation4.23}), and using the result (\ref{equation4.21})
gives
\begin{equation}
\partial_{n}M_{mj}^{i} = T_{k}^{i}R_{njm}^{k} + 2M_{kj}^{i}
\Gamma_{mn}^{k} -2 T_{l}\Gamma_{kj}^{l}\Gamma_{mn}^{k} \notag
\end{equation}
If this equation is integrated, the result will be
\begin{align}
M_{mj}^{i} = &M_{mj}^{i} (0) + 2M_{kj}^{i} (0) \int \Gamma_{mn}^{k}
dx^{n} + 4 M_{lj}^{i} (0) \int \Gamma_{mn}^{k} dx^{n} \int
\Gamma_{ka}^{l} dx^{a} \notag\\ &+ \int T_{k}^{i}(R^{k}_{njm} - 2
\Gamma_{lj}^{k} \Gamma_{mn}^{l}) dx^{n} + H.O. \notag
\end{align}
and using $T_{m}^{i} = T_{m}^{i} (0) + \int M_{mj}^{i} dx^{j}$ we
find $T$ up to the third order terms as follows
\begin{align}
T_{m}^{i} = &T_{m}^{i} (0) + M_{mj}^{i} (0) x^{j} + 2M_{kj}^{i} (0)
\int dx^{j} \int \Gamma_{mn}^{k} dx^{n} \notag\\ &+ 4 M_{lj}^{i} (0)
\int dx^{j} \int \Gamma_{mn}^{k} dx^{n} \int \Gamma_{ka}^{l} dx^{a}
+ T_{k}^{i} (0) \int dx^{j} \int (R^{k}_{njm} -
2\Gamma_{lj}^{k}\Gamma_{mn}^{l}) dx^{n} \notag\\ &+ M_{kb}^{i} (0)
\int dx^{j} \int (R^{k}_{njm} - 2\Gamma_{lj}^{k}
\Gamma_{mn}^{l})x^{b} dx^{n} + H.O. \notag
\end{align}
which immediately leads to a final result for $M_{mj}^{i}$
\begin{align}
M_{mj}^{i} &= M_{mj}^{i} (0) + 2M_{kj}^{i} (0) \int \Gamma_{mn}^{k}
dx^{n} + 4 M_{lj}^{i} (0) \int \Gamma_{mn}^{k} dx^{n} \int
\Gamma_{ka}^{l} dx^{a} \notag\\ &+ T_{k}^{i} (0)\int (R^{k}_{njm} -
2 \Gamma_{lj}^{k} \Gamma_{mn}^{l}) dx^{n} + M_{ka}^{i} (0)\int
(R^{k}_{njm} - 2 \Gamma_{lj}^{k} \Gamma_{mn}^{l}) x^{a} dx^{n} +
H.O. \notag
\end{align}
One may find torsion and curvature relations using these explicit
solutions as in the previous section. Let us inquire solutions by
expressing equations (\ref{equation4.13}) - (\ref{equation4.16}) in
terms of $T$ instead of finding explicit solutions.

If (\ref{equation4.21}) is inserted in the pseudoduality equations
(\ref{equation4.13})-(\ref{equation4.16}) we get
\begin{align} \label{equation4.35}
\tilde{\psi}_{+}^{i} &= + T_{j}^{i} \psi_{+}^{j} \\
\partial_{-}\tilde{\psi}_{+}^{i} &= - T_{j}^{i}
\partial_{-}\psi_{+}^{j} - M_{jm}^{i} \psi_{+}^{m}
\partial_{-}x^{j} \label{equation4.36}\\
\partial_{+}\tilde{x}^{i} &= + T_{j}^{i}
\partial_{+}x^{j} - iM_{jm}^{i}\psi_{+}^{m}\psi_{+}^{j}
\label{equation4.37}
\\
\partial_{-}\tilde{x}^{i} &= - T_{j}^{i}
\partial_{-}x^{j} \label{equation4.38}
\end{align}
Using equations of motion for $\partial_{-}\tilde{\psi}_{+}^{i}$ and
$\partial_{-}\psi_{+}^{j}$ in (\ref{equation4.36}), one finds
\begin{align} \label{equation4.39}
(\tilde{H}_{mn}^{i} -
\tilde{\Gamma}_{mn}^{i})\tilde{\psi}_{+}^{m}\partial_{-}\tilde{x}^{n}
= T_{j}^{i} \Gamma_{mn}^{j} \psi_{+}^{m} \partial_{-}x^{n} -
M_{nm}^{i} \psi_{+}^{m} \partial_{-}x^{n}
\end{align}
and inserting (\ref{equation4.35}) and (\ref{equation4.38}) into
(\ref{equation4.39}) leads to the following result
\begin{equation}
(\tilde{H}_{mn}^{i} - \tilde{\Gamma}_{mn}^{i})T_{a}^{m} T_{b}^{n} =
M_{ba}^{i} - T_{j}^{i} \Gamma_{ab}^{j} \label{equation4.40}
\end{equation}
Now taking $\partial_{-}$ of (\ref{equation4.37}) (or $\partial_{+}$
of \ref{equation4.38}) leads to
\begin{equation}
\partial^{2}_{+-}\tilde{x}^{i} =
M_{jk}^{i}\partial_{+}x^{j}\partial_{-}x^{k} +
T_{j}^{i}\partial^{2}_{+-}x^{j} -i\partial_{n}M_{jm}^{i}
\psi_{+}^{m}\psi_{+}^{j}\partial_{-}x^{n} - iM_{jm}^{i}
\partial_{-}\psi_{+}^{m}\psi_{+}^{j} - iM_{jm}^{i} \psi_{+}^{m}
\partial_{-}\psi_{+}^{j} \notag
\end{equation}
We use the equation of motion for $\partial^{2}_{+-}\tilde{x}^{i}$,
$\partial^{2}_{+-}x^{j}$ and $\partial_{-}\psi_{+}^{m}$, and use the
result (\ref{equation4.23}) to get
\begin{align}
(\tilde{H}_{jk}^{i} -
\tilde{\Gamma}_{jk}^{i})\partial_{+}\tilde{x}^{j}\partial_{-}\tilde{x}^{k}
+\frac{i}{2}\hat{\tilde{R}}^{i}_{jkm}\tilde{\psi}_{+}^{k}\tilde{\psi}_{+}^{m}\partial_{-}\tilde{x}^{j}
= &(M_{mn}^{i} -
T_{j}^{i}\Gamma_{mn}^{j})\partial_{+}x^{m}\partial_{-}x^{n}
\label{equation4.41}\\ &- \frac{i}{2} T_{j}^{i}
R^{j}_{mnk}\psi_{+}^{n}\psi_{+}^{k}\partial_{-}x^{m} \notag
\end{align}
now using (\ref{equation4.35}), (\ref{equation4.37}) and
(\ref{equation4.38}) in (\ref{equation4.41}) leads to
\begin{align}
&-(\tilde{H}_{jk}^{i} -
\tilde{\Gamma}_{jk}^{i})T_{m}^{j}T_{n}^{k}\partial_{+}x^{m}\partial_{-}x^{n}
+ i(\tilde{H}_{jl}^{i} -
\tilde{\Gamma}_{jl}^{i})M_{kn}^{j}T_{m}^{l}\psi_{+}^{n}\psi_{+}^{k}\partial_{-}x^{m}
\notag\\ &-
\frac{i}{2}\hat{\tilde{R}}^{i}_{abc}T_{n}^{b}T_{k}^{c}T_{m}^{a}\psi_{+}^{n}\psi_{+}^{k}\partial_{-}x^{m}
= (M_{mn}^{i} -
T_{j}^{i}\Gamma_{mn}^{j})\partial_{+}x^{m}\partial_{-}x^{n} -
\frac{i}{2} T_{j}^{i}
R^{j}_{mnk}\psi_{+}^{n}\psi_{+}^{k}\partial_{-}x^{m} \notag
\end{align}
which can be split into the following equations
\begin{align}
&(\tilde{H}_{jk}^{i} - \tilde{\Gamma}_{jk}^{i})T_{m}^{j}T_{n}^{k} =
-M_{mn}^{i} + T_{j}^{i}\Gamma_{mn}^{j} = M_{[nm]}^{i} \label{equation4.42} \\
&\frac{1}{2} T_{j}^{i} R^{j}_{mnk} =
\frac{1}{2}\hat{\tilde{R}}^{i}_{abc}T_{n}^{b}T_{k}^{c}T_{m}^{a} -
(\tilde{H}_{jl}^{i} - \tilde{\Gamma}_{jl}^{i})M_{[kn]}^{j}T_{m}^{l}
\label{equation4.43}
\end{align}
we see that (\ref{equation4.40}) and (\ref{equation4.42})  are the
same equations (by means of equation (\ref{equation4.21})). It is
evident that right hand side of equation (\ref{equation4.42}) is
equal to the antisymmetric part of $M_{nm}^{i}$, and therefore,
$\tilde{\Gamma}_{jk}^{i} = 0$. Equation (\ref{equation4.43}) can be
written as
\begin{equation}
\frac{1}{2} T_{j}^{i} R^{j}_{mnk} =
\frac{1}{2}(\hat{\tilde{R}}^{i}_{abc} -
2\tilde{H}_{ja}^{i}\tilde{H}_{bc}^{j})T_{n}^{b}T_{k}^{c}T_{m}^{a}
\label{equation4.44}
\end{equation}
where we used (\ref{equation4.42}). $\tilde{H}$ can be figured out
by (\ref{equation4.42}) using the initial values of $T$ and $M$,
hence it is easy to see that $\tilde{H}_{mn}^{i} = M_{[nm]}^{i}(0)$.
Therefore, we can write $\hat{\tilde{R}}^{i}_{abc} -
2\tilde{H}_{ja}^{i}\tilde{H}_{bc}^{j} = \tilde{R}^{i}_{abc}$, which
leads to $R^{i}_{mnk} = \tilde{R}^{i}_{mnk}$ by equation
(\ref{equation4.44}). This means that curvatures will be related to
each other by the relation $R^{i}_{mnk} = \tilde{R}^{i}_{mnk}$
around the point $p$ on $M$ where the transformation is identity,
and $\tilde{R}^{i}_{mnk}$ is the curvature at point $\tilde{p}$. In
this case all the points on manifold $M$ will be mapped to only one
point $\tilde{p}$ on $\tilde{M}$ where riemann normal coordinates
are used.

\subsection{Orthonormal Coframes} \label{sec4:OC}

In this case we will present pseudoduality equations on the
orthonormal coframe $SO(\mathbb{M})$. Equations of motion following
from the action (\ref{equation4.6}) in terms of the superfields are
\begin{equation}
X_{+-}^{k} = X_{-+}^{k} = - [\mathbf{\Gamma}_{ij}^{k}(X) -
\mathcal{H}_{ij}^{k}(X)]X_{+}^{i}X_{-}^{j} \label{equation4.45}
\end{equation}
where superfield $X$ has the form (\ref{equation4.5}), $D_{+}X =
X_{+}$ and $\partial_{-}X = X_{-}$. We choose an orthonormal frame
$\{\Lambda^{i}\}$ with the riemannian connection $\Lambda_{j}^{i}$
on the superspace. If the superspace coordinates are defined by $z =
(\sigma^{\pm}, \theta^{+})$, then one form is given by
\begin{equation}
\Lambda^{i} = dz^{M}X_{M}^{i} \label{equation4.46}
\end{equation}
Covariant derivatives of $X_{M}$ and $X_{MN}$ will be
\begin{equation}
dX_{M}^{i} + \Lambda_{j}^{i}X_{M}^{j} = dz^{N}X_{MN}^{i}
\label{equation4.47}
\end{equation}
The Cartan structural equations are
\begin{align}
d\Lambda^{i} &= - \Lambda_{j}^{i} \wedge \Lambda^{j}
\label{equation4.48}\\
d\Lambda_{j}^{i} &= -\Lambda_{k}^{i} \wedge \Lambda_{j}^{k} +
\Omega_{j}^{i} \label{equation4.49}
\end{align}
where $\Omega_{j}^{i} = \frac{1}{2}\mathcal{R}_{jkl}^{i}\Lambda^{k}
\wedge \Lambda^{l}$ is the curvature two form. Pseudoduality
equations (\ref{equation4.11}) and (\ref{equation4.12}) are
\begin{align}
\tilde{X}_{\pm}^{i} = \pm \mathcal{T}_{j}^{i} X_{\pm}^{j}
\label{equation4.50}
\end{align}
where $\mathcal{T}$ depends on superfield $X$. Taking the exterior
derivative of both sides yields
\begin{equation}
d\tilde{X}_{\pm}^{i} = \pm d\mathcal{T}_{j}^{i} X_{\pm}^{j} \pm
\mathcal{T}_{j}^{i} dX_{\pm}^{j} \notag
\end{equation}
Inserting (\ref{equation4.47}) in this equation gives
\begin{equation}
-\tilde{\Lambda}_{j}^{i} \tilde{X}_{\pm}^{j} + dz^{N}
\tilde{X}^{i}_{\pm N} = \pm d\mathcal{T}_{j}^{i} X_{\pm}^{j} \mp
\mathcal{T}_{j}^{i} \Lambda_{k}^{j} X_{\pm}^{k} \pm  dz^{N}
\mathcal{T}_{j}^{i}X_{\pm N}^{j} \notag
\end{equation}
We now substitute (\ref{equation4.50}) and arrange the terms to get
\begin{equation}
dz^{N} \tilde{X}^{i}_{\pm N} = \pm (d\mathcal{T}_{k}^{i} -
\mathcal{T}_{j}^{i}\Lambda_{k}^{j} +
\tilde{\Lambda}_{j}^{i}\mathcal{T}_{k}^{j})X_{\pm}^{k} \pm dz^{N}
\mathcal{T}_{j}^{i} X_{\pm N}^{j} \notag
\end{equation}
We wedge the plus equation (upper sign) by $dz^{+}$ and minus
equation (lower sign) by $dz^{-}$, and find the following equations
\begin{align}
dz^{+} \wedge dz^{-} \tilde{X}_{+-}^{i} = dz^{+} \wedge
(d\mathcal{T}_{k}^{i} - \mathcal{T}_{j}^{i}\Lambda_{k}^{j} +
\tilde{\Lambda}_{j}^{i}\mathcal{T}_{k}^{j})X_{+}^{k} + dz^{+} \wedge
dz^{-}
\mathcal{T}_{j}^{i} X_{+-}^{j} \label{equation4.51}\\
dz^{-} \wedge dz^{+} \tilde{X}_{-+}^{i} = - dz^{-} \wedge
(d\mathcal{T}_{k}^{i} - \mathcal{T}_{j}^{i}\Lambda_{k}^{j} +
\tilde{\Lambda}_{j}^{i}\mathcal{T}_{k}^{j})X_{-}^{k} - dz^{-} \wedge
dz^{+} \mathcal{T}_{j}^{i} X_{-+}^{j} \label{equation4.52}
\end{align}
Since $X_{+-} = X_{-+}$ (also with tilde) and $dz^{+} \wedge dz^{-}
= dz^{-} \wedge dz^{+}$ we may find the constraint relations by
equating left hand sides
\begin{align}
2 dz^{+} \wedge dz^{-} \mathcal{T}_{k}^{i} X_{+-}^{k} + dz^{+}
\wedge (d\mathcal{T}_{k}^{i} - \mathcal{T}_{j}^{i}\Lambda_{k}^{j} +
\tilde{\Lambda}_{j}^{i}\mathcal{T}_{k}^{j})X_{+}^{k} \notag\\ +
dz^{-} \wedge (d\mathcal{T}_{k}^{i} -
\mathcal{T}_{j}^{i}\Lambda_{k}^{j} +
\tilde{\Lambda}_{j}^{i}\mathcal{T}_{k}^{j})X_{-}^{k} = 0
\label{equation4.53}
\end{align}
we substitute the equations of motion (\ref{equation4.45})
\begin{align}
&- 2 dz^{+} \wedge dz^{-} \mathcal{T}_{k}^{i}
[\mathbf{\Gamma}_{mn}^{k} - \mathcal{H}_{mn}^{k}]X_{+}^{m}X_{-}^{n}
+ dz^{+} \wedge (d\mathcal{T}_{k}^{i} -
\mathcal{T}_{j}^{i}\Lambda_{k}^{j} +
\tilde{\Lambda}_{j}^{i}\mathcal{T}_{k}^{j})X_{+}^{k} \notag\\ &+
dz^{-} \wedge (d\mathcal{T}_{k}^{i} -
\mathcal{T}_{j}^{i}\Lambda_{k}^{j} +
\tilde{\Lambda}_{j}^{i}\mathcal{T}_{k}^{j})X_{-}^{k} = 0
\label{equation4.54}
\end{align}
and we use $dz^{\pm} X_{\pm}^{n} = \Lambda^{n} -
dz^{\mp}X_{\mp}^{n}$ to get
\begin{align}
&-dz^{+} \wedge \mathcal{T}_{k}^{i} (\mathbf{\Gamma}_{mn}^{k} -
\mathcal{H}_{mn}^{k})X_{+}^{m}\Lambda^{n} - dz^{-} \wedge
\mathcal{T}_{k}^{i} (\mathbf{\Gamma}_{mn}^{k} +
\mathcal{H}_{mn}^{k})X_{-}^{m}\Lambda^{n} \label{equation4.55}\\ &+
dz^{+} \wedge (d\mathcal{T}_{k}^{i} -
\mathcal{T}_{j}^{i}\Lambda_{k}^{j} +
\tilde{\Lambda}_{j}^{i}\mathcal{T}_{k}^{j})X_{+}^{k} + dz^{-} \wedge
(d\mathcal{T}_{k}^{i} - \mathcal{T}_{j}^{i}\Lambda_{k}^{j} +
\tilde{\Lambda}_{j}^{i}\mathcal{T}_{k}^{j})X_{-}^{k} = 0 \notag
\end{align}
Now we define the following tensors
\begin{align}
dz^{-}\mathcal{U}_{k-}^{i} &= (d\mathcal{T}_{k}^{i} -
\mathcal{T}_{j}^{i}\Lambda_{k}^{j} +
\tilde{\Lambda}_{j}^{i}\mathcal{T}_{k}^{j}) -
\mathcal{T}_{j}^{i}(\mathbf{\Gamma}_{kn}^{j} -
\mathcal{H}_{kn}^{j})\Lambda^{n} \label{equation4.56}\\
dz^{+}\mathcal{U}_{k+}^{i} &= - (d\mathcal{T}_{k}^{i} -
\mathcal{T}_{j}^{i}\Lambda_{k}^{j} +
\tilde{\Lambda}_{j}^{i}\mathcal{T}_{k}^{j})
+\mathcal{T}_{j}^{i}(\mathbf{\Gamma}_{kn}^{j} +
\mathcal{H}_{kn}^{j})\Lambda^{n} \label{equation4.57}
\end{align}
which satisfies the equation (\ref{equation4.55})
\begin{equation}
dz^{+} \wedge dz^{-} \mathcal{U}_{k-}^{i}X_{+}^{k} - dz^{-} \wedge
dz^{+} \mathcal{U}_{k+}^{i}X_{-}^{k} = 0 \label{equation4.58}
\end{equation}
They also yield the result
\begin{equation}
dz^{-}\mathcal{U}_{k-}^{i} + dz^{+}\mathcal{U}_{k+}^{i} =
2\mathcal{T}_{j}^{i}\mathcal{H}_{kn}^{j}\Lambda^{n}
\label{equation4.59}
\end{equation}
which gives
\begin{align}
dz^{+} \wedge dz^{-} \mathcal{U}_{k-}^{i} = 2dz^{+} \wedge
\mathcal{T}_{j}^{i}\mathcal{H}_{kn}^{j}\Lambda^{n} \label{equation4.60}\\
dz^{-} \wedge dz^{+} \mathcal{U}_{k+}^{i} = 2dz^{-} \wedge
\mathcal{T}_{j}^{i}\mathcal{H}_{kn}^{j}\Lambda^{n}
\label{equation4.61}
\end{align}
If these equations are substituted into (\ref{equation4.58}), one
obtains
\begin{equation}
2dz^{+} \wedge
\mathcal{T}_{j}^{i}\mathcal{H}_{kn}^{j}\Lambda^{n}X_{+}^{k} -
2dz^{-} \wedge
\mathcal{T}_{j}^{i}\mathcal{H}_{kn}^{j}\Lambda^{n}X_{-}^{k} = 0
\label{equation4.62}
\end{equation}
and using (\ref{equation4.46}) gives the final result
\begin{equation}
2dz^{+} \wedge dz^{-}
\mathcal{T}_{j}^{i}\mathcal{H}_{kn}^{j}X_{-}^{n}X_{+}^{k} - 2dz^{-}
\wedge dz^{+}
\mathcal{T}_{j}^{i}\mathcal{H}_{kn}^{j}X_{+}^{n}X_{-}^{k} = 0
\label{equation4.63}
\end{equation}
which shows that
\begin{equation}
dz^{+} \wedge dz^{-}
\mathcal{T}_{j}^{i}\mathcal{H}_{kn}^{j}X_{+}^{k}X_{-}^{n} = 0
\label{equation4.64}
\end{equation}

Therefore, we conclude that $\mathcal{H} = 0$, and
$\mathcal{U}_{k-}^{i} = \mathcal{U}_{k+}^{i} = 0$ by equations
(\ref{equation4.60}) and (\ref{equation4.61}). Finally equation
(\ref{equation4.56}) and (\ref{equation4.57}) gives the following
result
\begin{equation}
(d\mathcal{T}_{k}^{i} - \mathcal{T}_{j}^{i}\Lambda_{k}^{j} +
\tilde{\Lambda}_{j}^{i}\mathcal{T}_{k}^{j}) =
\mathcal{T}_{j}^{i}\mathbf{\Gamma}_{kn}^{j}\Lambda^{n}
\label{equation4.65}
\end{equation}

If we insert the equations of motion into (\ref{equation4.51}) and
(\ref{equation4.52}), we obtain
\begin{align}
-dz^{+} \wedge dz^{-} (\mathbf{\tilde{\Gamma}}_{jk}^{i} -
\mathcal{\tilde{H}}_{jk}^{i})\tilde{X}_{+}^{j}\tilde{X}_{-}^{k} =
dz^{+} \wedge (d\mathcal{T}_{k}^{i} -
\mathcal{T}_{j}^{i}\Lambda_{k}^{j} +
\tilde{\Lambda}_{j}^{i}\mathcal{T}_{k}^{j})X_{+}^{k} \notag\\ -
dz^{+} \wedge
dz^{-} \mathcal{T}_{j}^{i} \mathbf{\Gamma}_{mn}^{i} X_{+}^{m}X_{-}^{n} \label{equation4.66}\\
-dz^{-} \wedge dz^{+} (\mathbf{\tilde{\Gamma}}_{jk}^{i} -
\mathcal{\tilde{H}}_{jk}^{i})\tilde{X}_{+}^{j}\tilde{X}_{-}^{k} = -
dz^{-} \wedge (d\mathcal{T}_{k}^{i} -
\mathcal{T}_{j}^{i}\Lambda_{k}^{j} +
\tilde{\Lambda}_{j}^{i}\mathcal{T}_{k}^{j})X_{-}^{k} \notag\\ +
dz^{-} \wedge dz^{+} \mathcal{T}_{j}^{i} \mathbf{\Gamma}_{mn}^{i}
X_{+}^{m}X_{-}^{n} \label{equation4.67}
\end{align}
Inserting $dz^{-}X_{-}^{k} = \Lambda - dz^{+}X_{+}$ (also with
tilde) and $\tilde{X}_{+}$ (\ref{equation4.50}) into
(\ref{equation4.66}), and $dz^{+}X_{+} = \Lambda - dz^{-}X_{-}$
(also with tilde) and $\tilde{X}_{-}$ (\ref{equation4.50}) into
(\ref{equation4.67}) gives
\begin{align}
-dz^{+} \wedge (\mathbf{\tilde{\Gamma}}_{mn}^{i} -
\mathcal{\tilde{H}}_{mn}^{i})T_{k}^{m} \tilde{\Lambda}^{n} = dz^{+}
\wedge (d\mathcal{T}_{k}^{i} - \mathcal{T}_{j}^{i}\Lambda_{k}^{j} +
\tilde{\Lambda}_{j}^{i}\mathcal{T}_{k}^{j}) - dz^{+} \wedge
\mathcal{T}_{j}^{i}
\Gamma_{kn}^{i} \Lambda^{n} \label{equation4.68}\\
- dz^{-} \wedge (\mathbf{\tilde{\Gamma}}_{mn}^{i} +
\mathcal{\tilde{H}}_{mn}^{i})T_{k}^{m} \tilde{\Lambda}^{n} = -
dz^{-} \wedge (d\mathcal{T}_{k}^{i} -
\mathcal{T}_{j}^{i}\Lambda_{k}^{j} +
\tilde{\Lambda}_{j}^{i}\mathcal{T}_{k}^{j}) + dz^{-} \wedge
\mathcal{T}_{j}^{i} \Gamma_{kn}^{i} \Lambda^{n} \label{equation4.69}
\end{align}
where we cancelled out $X_{+}^{k}$ in (\ref{equation4.68}) and
$X_{-}^{k}$ in (\ref{equation4.69}). We notice that right-hand sides
of these equations become zero by means of the constraint relation
(\ref{equation4.65}), and we are left with
\begin{align}
(\mathbf{\tilde{\Gamma}}_{mn}^{i} -
\mathcal{\tilde{H}}_{mn}^{i})\mathcal{T}_{k}^{m} \tilde{\Lambda}^{n} = 0 \label{equation4.70}\\
(\mathbf{\tilde{\Gamma}}_{mn}^{i} +
\mathcal{\tilde{H}}_{mn}^{i})\mathcal{T}_{k}^{m} \tilde{\Lambda}^{n}
= 0 \label{equation4.71}
\end{align}
This shows that on the transformed superspace we must have
$\mathbf{\tilde{\Gamma}} = 0$ and $\mathcal{\tilde{H}}_{mn}^{i} =
0$. We may find the relation between curvatures of the spaces using
(\ref{equation4.65}). We  may define the connection one form
$\Lambda_{k}^{j} = \mathbf{\Gamma}_{kn}^{j} \Lambda^{n}$, and hence
(\ref{equation4.65}) is reduced to
\begin{equation}
(d\mathcal{T}_{k}^{i} - 2 \mathcal{T}_{j}^{i}\Lambda_{k}^{j} +
\tilde{\Lambda}_{j}^{i}\mathcal{T}_{k}^{j}) = 0 \label{equation4.72}
\end{equation}
Taking exterior derivative, and using again (\ref{equation4.72})
together with (\ref{equation4.49}) gives
\begin{equation}
\mathcal{T}_{j}^{i}\Omega_{k}^{j} = \tilde{\Omega}_{j}^{i}
\mathcal{T}_{k}^{j}\label{equation4.73}
\end{equation}
where new orthonormal coframe is replaced by $2\Lambda$ with the
same curvature two form $\Omega$ on the manifold $\mathbb{M}$. It is
obvious that integrability condition of this equation followed by
the use of (\ref{equation4.46}) and (\ref{equation4.50}) yields a
curvature relation between two (1, 0) supersymmetric sigma models
which tied together with pseudoduality, which can be reduced to the
same results found in the previous section. The reason why we get a
positive sign in curvature expression in component expansion method
is because of anticommuting grassmann numbers. This gives that
pseudoduality transformation can be performed only if two sigma
models are based on symmetric spaces with opposite curvatures on
target spaces $\mathbb{M}$ and $\tilde{\mathbb{M}}$.

\section{Pseudoduality in (1, 1) Supersymmetric Sigma
Models}\label{sec4:PSSM}

In this case \cite{ketov1} the classical spacetime $\Sigma$ can be
enlarged to the superspace $\Xi^{1, 1}$ by adding Grassmann
coordinates of opposite chirality. We will have one left-handed
supercharge $Q_{+}$, and one right-handed supercharge $Q_{-}$ as
given by (\ref{equation4.2}). The supersymmetry algebra can be
written as
\begin{equation}
\{Q_{\pm}, Q_{\pm}\} = 2 i P_{\pm} \ \ \ \ \ \ \ \ \ \ \{Q_{+},
Q_{-}\} = 0 \notag
\end{equation}
where $P_{\pm} = - \partial_{\pm}$. The supersymmetry
transformations will be
\begin{align}
\delta_{\epsilon} x^{i} &= \epsilon^{+} \psi_{+}^{i} + \epsilon^{-}
\psi_{-}^{i}
\notag\\
\delta_{\epsilon} \psi_{+}^{i} &= i \epsilon^{+} \partial_{+} x^{i}
+ \epsilon^{-} (\Gamma_{j k}^{i} + H_{j k}^{i}) \psi_{+}^{i}
\psi_{-}^{k} \notag\\
\delta_{\epsilon} \psi_{-}^{i} &= - \epsilon^{+} (\Gamma_{j k}^{i} +
H_{j k}^{i}) \psi_{+}^{j} \psi_{-}^{k} + i \epsilon^{-}
\partial_{-}x^{i} \notag
\end{align}
where $\epsilon^{\pm}$ is the constant anticommuting parameter.

\subsection{Components}

The superfield is written as
\begin{equation}
X = x + \theta^{+}\psi_{+} + \theta^{-}\psi_{-} +
\theta^{+}\theta^{-}F \label{equation4.74}
\end{equation}
where  $X :\Xi^{1, 1} \rightarrow M$. We define the $(1, 1)$
superspace $\Xi^{1, 1} = (\sigma^{+}, \sigma^{-}, \theta^{+},
\theta^{-})$, where $(\sigma^{+}, \sigma^{-})$ are the null
coordinates, and $(\theta^{+}, \theta^{-})$ are the Grassman
coordinates of opposite chirality. The action of the theory is
\begin{equation}
S = \int d^{2}\sigma d^{2}\theta (G_{ij} + B_{ij}) D_{+}X^{i}
D_{-}X^{j} \label{equation4.75}
\end{equation}
where supercovariant derivatives are given in (\ref{equation4.1}).
Similar definitions can be written for pseudodual model with tilde.
First order expansion of $G_{ij}$ and $B_{ij}$, followed by the
$d^{2}\theta$ integral gives
\begin{equation}
S = - \int d^{2}x [(g_{ij} + b_{ij})\partial_{+}x^{i}
\partial_{-}x^{j} + ig_{ij}\psi_{+}^{i} \nabla_{-}^{(-)}\psi_{+}^{j} + ig_{ij}\psi_{-}^{i}\nabla_{+}^{(+)}\psi_{-}^{j} -
\frac{1}{2}\hat{R}^{+}_{bnam}\psi_{+}^{m}\psi_{-}^{n}\psi_{+}^{a}\psi_{-}^{b}]
\notag
\end{equation}
where $\nabla_{\pm}^{(\pm)}\psi_\mp^{j} = \nabla_{\pm}\psi_{\mp}^{j}
\pm H_{mn}^{j}\psi_{\mp}^{m} \partial_{\pm}x^{n}$, and
$\hat{R}^{\pm}_{bnam} = R_{bnam} \pm D_{a}H_{nmb} \mp D_{m}H_{nab} +
H_{baj}H^{j}_{mn} - H_{naj}H^{j}_{mb}$.

Equations of motion following from this action will be
\begin{align} \label{equation4.76}
F^{i} = &(\Gamma_{jk}^{i} - H_{jk}^{i}) \psi_{+}^{j} \psi_{-}^{k} \\
\nabla_{-}^{(-)}\psi_{+}^{i} = &\frac{i}{2} (\hat{R}^{+})^{i}_{jmn}
\psi_{-}^{n} \psi_{+}^{j} \psi_{-}^{m} \label{equation4.77}\\
\nabla_{+}^{(+)}\psi_{-}^{i} = &\frac{i}{2} (\hat{R}^{+})^{i}_{jmn}
\psi_{+}^{n} \psi_{-}^{j} \psi_{+}^{m} \label{equation4.78} \\
\square x^{k} = &i (\hat{R}^{-})^{k}_{nim} \psi_{+}^{i} \psi_{+}^{m}
\partial_{-}x^{n} + i (\hat{R}^{+})^{k}_{nim} \psi_{-}^{i}
\psi_{-}^{m} \partial_{+}x^{n} \label{equation4.79}\\ &-
(\hat{D}^{k}\hat{R}^{+}_{bnam}) \psi_{+}^{m} \psi_{-}^{n}
\psi_{+}^{a} \psi_{-}^{b} \notag
\end{align}
where $\hat{D}^{k}\hat{R}^{+}_{bnam} = D^{k}\hat{R}^{+}_{bnam} +
H^{k}_{jn}(\hat{R}^{+})^{j}_{bam} -
H^{k}_{jb}(\hat{R}^{+})^{j}_{nam} +
H^{k}_{ja}(\hat{R}^{+})^{j}_{mbn} -
H^{k}_{jm}(\hat{R}^{+})^{j}_{abn}$.

Pseudoduality transformations are
\begin{align} \label{equation4.80}
D_{+}\tilde{X}^{i} &= + \mathcal{T}_{j}^{i} D_{+}X^{j} \\
D_{-}\tilde{X}^{i} &= - \mathcal{T}_{j}^{i} D_{-}X^{j}
\label{equation4.81}
\end{align}
where $\mathcal{T}$ is a function of superfield
(\ref{equation4.74}). Transformation matrix $\mathcal{T}$ can be
expanded as $\mathcal{T}(X) = T(x) +
\theta^{+}\psi_{+}^{k}\partial_{k}T(x) +
\theta^{-}\psi_{-}^{k}\partial_{k}T(x) +
\theta^{+}\theta^{-}F^{k}\partial_{k}T(x) -
\theta^{+}\theta^{-}\psi_{+}^{k}\psi_{-}^{l}\partial_{k}\partial_{l}T(x)$.
If pseudoduality transformations are written in components, first
equation (\ref{equation4.80}) yields the following set of equations
\begin{align} \label{equation4.82}
\tilde{\psi}_{+}^{i} &= T_{j}^{i} \psi_{+}^{j} \\
\tilde{F}^{i} &= T_{j}^{i} F^{j} - M_{jk}^{i}\psi_{+}^{j}\psi_{-}^{k} \label{equation4.83}\\
\partial_{+}\tilde{x}^{i} &= T_{j}^{i}\partial_{+}x^{j} +
iM_{jk}^{i}\psi_{+}^{j}\psi_{+}^{k} \label{equation4.84}\\
\partial_{+}\tilde{\psi}_{-}^{i} &=
T_{j}^{i}\partial_{+}\psi_{-}^{j} -2iM_{[jk]}^{i}\psi_{+}^{j}F^{k} +
M_{kj}^{i}\psi_{-}^{j}\partial_{+}x^{k} +
i\partial_{l}M_{[jk]}^{i}\psi_{+}^{k}\psi_{-}^{l}\psi_{+}^{j}
\label{equation4.85}
\end{align}
where $M_{jk}^{i} = \partial_{k}T_{j}^{i}$, and $M_{[jk]}^{i}$
represents the antisymmetric part of $M_{jk}^{i}$. Second equation
(\ref{equation4.81}) will produce
\begin{align} \label{equation4.86}
\tilde{\psi}_{-}^{i} &= -T_{j}^{i} \psi_{-}^{j} \\
\tilde{F}^{i} &= -T_{j}^{i} F^{j} + M_{kj}^{i}\psi_{+}^{j}\psi_{-}^{k} \label{equation4.87}\\
\partial_{-}\tilde{x}^{i} &= -T_{j}^{i}\partial_{-}x^{j} -
iM_{jk}^{i}\psi_{-}^{j}\psi_{-}^{k} \label{equation4.88}\\
\partial_{-}\tilde{\psi}_{+}^{i} &=
-T_{j}^{i}\partial_{-}\psi_{+}^{j} -2iM_{[jk]}^{i}\psi_{-}^{j}F^{k}
- M_{kj}^{i}\psi_{+}^{j}\partial_{-}x^{k} -
i\partial_{l}M_{[jk]}^{i}\psi_{-}^{k}\psi_{+}^{l}\psi_{-}^{j}
\label{equation4.89}
\end{align}

We can find constraint relations using these equations. If
(\ref{equation4.83}) is set equal to (\ref{equation4.87}), and
equation of motion (\ref{equation4.76}) is used, the result follows
\begin{equation}
T_{j}^{i}(\Gamma_{mn}^{i} - H_{mn}^{j}) = M_{(mn)}^{i}
\label{equation4.90}
\end{equation}
where $M_{(mn)}^{i}$ is the symmetric part of $M_{mn}^{i}$. We
immediately notice that $H_{mn}^{j} = 0$, and we are left with
\begin{equation}
T_{j}^{i}\Gamma_{mn}^{j} = M_{(mn)}^{i} \label{equation4.91}
\end{equation}

 We next take $\partial_{-}$ of (\ref{equation4.82})and set equal to (\ref{equation4.89})
followed by the equations of motion (\ref{equation4.76}) and
(\ref{equation4.77}) to obtain
\begin{align} \label{equation4.92}
[2M_{(mn)}^{i}
-2T_{j}^{i}\Gamma_{mn}^{j}]\psi_{+}^{m}\partial_{-}x^{n} =
-i[\partial_{a}M_{[bc]}^{i} + 2M_{[ck]}^{i}\Gamma_{ab}^{k} +
T_{j}^{i}R^{j}_{abc}]\psi_{-}^{c}\psi_{+}^{a}\psi_{-}^{b}
\end{align}
Real part of this equation is simply (\ref{equation4.91}), and
complex part will produce
\begin{equation}
\partial_{a}M_{[bc]}^{i} + 2M_{[ck]}^{i}\Gamma_{ab}^{k} +
T_{j}^{i}R^{j}_{abc} = 0 \label{equation4.93}
\end{equation}

We now take $\partial_{+}$ of (\ref{equation4.86}) and set equal to
(\ref{equation4.85}) followed by the equations of motion
(\ref{equation4.76}) and (\ref{equation4.78}) to get
\begin{equation}
[2M_{(mn)}^{i} -
2T_{j}^{i}\Gamma_{mn}^{j}]\psi_{-}^{m}\partial_{+}x^{n} =
-i[\partial_{a}M_{[bc]}^{i} + 2M_{[ck]}^{i}\Gamma_{ab}^{k} +
T_{j}^{i}R^{j}_{abc}]\psi_{+}^{c}\psi_{-}^{a}\psi_{+}^{b}
\label{equation4.94}
\end{equation}
This equation is similar to (\ref{equation4.92}), and we again
notice that real part of this equation is equal to
(\ref{equation4.91}), and complex part is (\ref{equation4.93}). We
finally take $\partial_{-}$ of (\ref{equation4.84}), $\partial_{+}$
of (\ref{equation4.88}), and set them equal to each other to find
out the remaining constraints
\begin{align}
2M_{(jk)}^{i} \partial_{+}x^{j}\partial_{-}x^{k} &+
2T_{j}^{i}\partial^{2}_{+-}x^{j} =
2iM_{[kj]}^{i}\psi_{+}^{j}\partial_{-}\psi_{+}^{k} +
i\partial_{n}M_{[kj]}\psi_{+}^{j}\psi_{+}^{k}\partial_{-}x^{n}
\notag\\ &+ 2iM_{[kj]}^{i}\psi_{-}^{j}\partial_{+}\psi_{-}^{k} +
i\partial_{n}M_{[kj]}\psi_{-}^{j}\psi_{-}^{k}\partial_{+}x^{n}
\label{equation4.95}
\end{align}
using equations of motion for $\partial^{2}_{+-}x^{j}$
(\ref{equation4.79}), $\partial_{-}\psi_{+}^{j}$
(\ref{equation4.77}) and $\partial_{+}\psi_{-}^{j}$
(\ref{equation4.78}) yields
\begin{align} \label{equation4.96}
&(2M_{(mn)}^{i} -
2T_{j}^{i}\Gamma_{mn}^{j})\partial_{+}x^{m}\partial_{-}x^{n} +
i(T_{j}^{i}R^{j}_{abc} + 2M_{[kb]}^{i}\Gamma_{ca}^{k} +
\partial_{a}M_{[bc]}^{i})\psi_{+}^{b}\psi_{+}^{c}\partial_{-}x^{a} \notag\\
&+ i(T_{j}^{i}R^{j}_{abc} + 2M_{[kb]}^{i}\Gamma_{ca}^{k} +
\partial_{a}M_{[bc]}^{i})\psi_{-}^{b}\psi_{-}^{c}\partial_{+}x^{a}
\notag\\ &- (T_{j}^{i}D^{j}R_{abcd} + M_{[dk]}^{i}R^{k}_{cab} +
M_{[bk]}^{i}R^{k}_{acd})\psi_{+}^{d}\psi_{-}^{b}\psi_{+}^{c}\psi_{-}^{a}
= 0
\end{align}
If this equation is split into real and complex parts the following
results are obtained
\begin{align}
(2M_{(mn)}^{i} -
2T_{j}^{i}\Gamma_{mn}^{j})\partial_{+}x^{m}\partial_{-}x^{n} =
(T_{j}^{i}D^{j}R_{abcd} + M_{[dk]}^{i}R^{k}_{cab} +
M_{[bk]}^{i}R^{k}_{acd})\psi_{+}^{d}\psi_{-}^{b}\psi_{+}^{c}\psi_{-}^{a}
\notag
\end{align}
\begin{align}
&(T_{j}^{i}R^{j}_{abc} + 2M_{[kb]}^{i}\Gamma_{ca}^{k} +
\partial_{a}M_{[bc]}^{i})\psi_{+}^{b}\psi_{+}^{c}\partial_{-}x^{a} \notag\\ &+ (T_{j}^{i}R^{j}_{abc} + 2M_{[kb]}^{i}\Gamma_{ca}^{k} +
\partial_{a}M_{[bc]}^{i})\psi_{-}^{b}\psi_{-}^{c}\partial_{+}x^{a} =
0 \notag
\end{align}
First equation leads to the following results
\begin{align}
&M_{(mn)}^{i} = T_{j}^{i}\Gamma_{mn}^{j} \label{equation4.97}\\
T_{j}^{i}D^{j}R_{abcd} &+ M_{[dk]}^{i}R^{k}_{cab} +
M_{[bk]}^{i}R^{k}_{acd} = 0 \label{equation4.98}
\end{align}
where (\ref{equation4.97}) is the same as (\ref{equation4.91}).
Second equation gives
\begin{equation}
T_{j}^{i}R^{j}_{abc} + 2M_{[kb]}^{i}\Gamma_{ca}^{k} +
\partial_{a}M_{[bc]}^{i} = 0 \label{equation4.99}
\end{equation}
which is the same equation as (\ref{equation4.93}) with
$b\leftrightarrow c$. Obviously we have three independent constraint
relations, which are (\ref{equation4.91}), (\ref{equation4.93}), and
(\ref{equation4.98}).

Now we can find out pseudodual fields, and relations between two
sigma models based on $M$ and $\tilde{M}$ by means of pseudoduality
equations. Using (\ref{equation4.83}) or (\ref{equation4.87}), and
equation of motion (\ref{equation4.76}) for $F^{j}$ we get
\begin{equation}
\tilde{F}^{i} = M_{[nm]}^{i}\psi_{+}^{m}\psi_{-}^{n}
\label{equation4.100}
\end{equation}
Also definition of $\tilde{F}^{i}$ gives that
\begin{align}
\tilde{F}^{i} &= (\tilde{\Gamma}_{jk}^{i} -
\tilde{H}_{jk}^{i})\tilde{\psi}_{+}^{j}\tilde{\psi}_{-}^{k} \notag\\
&= - (\tilde{\Gamma}_{jk}^{i} -
\tilde{H}_{jk}^{i})T_{m}^{j}T_{n}^{k}\psi_{+}^{m}\psi_{-}^{n}
\label{equation4.101}
\end{align}
where we used (\ref{equation4.82}) and (\ref{equation4.86}).
Comparison of (\ref{equation4.100}) with (\ref{equation4.101}) gives
that
\begin{equation}
(\tilde{\Gamma}_{jk}^{i} - \tilde{H}_{jk}^{i})T_{m}^{j}T_{n}^{k} =
M_{[mn]}^{i} \label{equation4.102}
\end{equation}

Hence we obtain that $\tilde{\Gamma}_{jk}^{i} = 0$. This means that
pseudoduality transformation will be from any point on $M$ to only
one point where $\tilde{\Gamma}$ vanishes on $\tilde{M}$. We know
that this is consistent with Riemann normal coordinates. We are left
with
\begin{equation}
\tilde{H}_{jk}^{i}T_{m}^{j}T_{n}^{k} = M_{[nm]}^{i}
\label{equation4.103}
\end{equation}

We next consider (\ref{equation4.85}). Using equations of motion
(\ref{equation4.76}) and (\ref{equation4.78}) we obtain
\begin{equation}
\partial_{+}\tilde{\psi}_{-}^{i} =
M_{[mn]}^{i}\psi_{-}^{m}\partial_{+}x^{n} -
\frac{i}{2}T_{j}^{i}R^{j}_{abc}\psi_{+}^{c}\psi_{-}^{a}\psi_{+}^{b}
\label{equation4.104}
\end{equation}
where we used the constraint (\ref{equation4.93}). On the other hand
we can write the equation of motion (\ref{equation4.78}) on
$\tilde{M}$ as
\begin{align}
\partial_{+}\tilde{\psi}_{-}^{i} &= -
\tilde{H}_{jk}^{i}\tilde{\psi}_{-}^{j}\partial_{+}\tilde{x}^{k} +
\frac{i}{2}(\hat{\tilde{R}}^{+})^{i}_{jmn} \tilde{\psi}_{+}^{n}
\tilde{\psi}_{-}^{j}\tilde{\psi}_{+}^{m} \label{equation4.105}\\ &=
\tilde{H}_{jk}^{i} T_{m}^{j}T_{n}^{k} \psi_{-}^{m} \partial_{+}x^{n}
+ i(\tilde{H}_{jk}^{i}T_{a}^{j}M_{[bc]}^{k} -
\frac{1}{2}(\hat{\tilde{R}}^{+})^{i}_{jmn}T_{c}^{n} T_{a}^{j}
T_{b}^{m})\psi_{+}^{c}\psi_{-}^{a}\psi_{+}^{b} \notag
\end{align}
where we used (\ref{equation4.82}), (\ref{equation4.84}) and
(\ref{equation4.86}) in the first line of (\ref{equation4.105}). If
we compare (\ref{equation4.104}) with (\ref{equation4.105}) we see
that
\begin{align}
\tilde{H}_{jk}^{i} T_{m}^{j}T_{n}^{k} &= M_{[mn]}^{i}
\label{equation4.106}\\
\frac{1}{2}T_{j}^{i}R^{j}_{abc} &=
\frac{1}{2}(\hat{\tilde{R}}^{+})^{i}_{jmn}T_{c}^{n} T_{a}^{j}
T_{b}^{m} - \tilde{H}_{jk}^{i}T_{a}^{j}M_{[bc]}^{k}
\label{equation4.107}
\end{align}

From (\ref{equation4.103}) and (\ref{equation4.106}) it is obvious
that antisymmetric part of $M_{mn}^{i}$ disappears, $M_{[mn]}^{i} =
0$, which leads to the result $\tilde{H}_{jk}^{i} = 0$. Hence
(\ref{equation4.107}) is reduced to
\begin{equation}
T_{j}^{i}R^{j}_{abc} = \tilde{R}^{i}_{jmn}T_{c}^{n} T_{a}^{j}
T_{b}^{m} \label{equation4.108}
\end{equation}

We now simplify right hand side of (\ref{equation4.89}). We use
equations of motion (\ref{equation4.76}) and (\ref{equation4.77})
and arrange the terms to get
\begin{equation}
\partial_{-}\tilde{\psi}_{+}^{i} = \frac{i}{2}T_{j}^{i}R^{j}_{abc} \psi_{-}^{c}\psi_{+}^{a}\psi_{-}^{b}
\label{equation4.109}
\end{equation}
where we used the constraint (\ref{equation4.93}). Also equation of
motion for $\partial_{-}\tilde{\psi}_{+}^{i}$ on $\tilde{M}$ gives
\begin{align}
\partial_{-}\tilde{\psi}_{+}^{i} &=
\frac{i}{2}\tilde{R}^{i}_{jmn}\tilde{\psi}_{-}^{n}\tilde{\psi}_{+}^{j}\tilde{\psi}_{-}^{m}
\notag\\
&=\frac{i}{2}\tilde{R}^{i}_{jmn}T_{a}^{n}T_{b}^{j}T_{c}^{m}\psi_{-}^{a}\psi_{+}^{b}\psi_{-}^{c}
\label{equation4.110}
\end{align}
A comparison of (\ref{equation4.109}) with (\ref{equation4.110})
gives (\ref{equation4.108}). When we take $\partial_{-}$ of
{\ref{equation4.84}}, and using relevant equations of motion
together with the constraints (\ref{equation4.93}) and
(\ref{equation4.98}) gives
\begin{equation}
\partial^{2}_{+-}\tilde{x}^{i} =
\frac{i}{2}T_{j}^{i}R^{j}_{abc}\psi_{-}^{b}\psi_{-}^{c}\partial_{+}x^{a}
-
\frac{i}{2}T_{j}^{i}R^{j}_{abc}\psi_{+}^{b}\psi_{+}^{c}\partial_{-}x^{a}
\label{equation4.111}
\end{equation}
Likewise on $\tilde{M}$ we obtain
\begin{align}
\partial^{2}_{+-}\tilde{x}^{i} =
&\frac{i}{2}\tilde{R}^{i}_{abc}\tilde{\psi}_{+}^{b}\tilde{\psi}_{+}^{c}\partial_{-}\tilde{x}^{a}
+
\frac{i}{2}\tilde{R}^{i}_{abc}\tilde{\psi}_{-}^{b}\tilde{\psi}_{-}^{c}\partial_{+}\tilde{x}^{a}
- \frac{1}{2}
\tilde{D}^{i}\tilde{R}_{abcd}\tilde{\psi}_{+}^{d}\tilde{\psi}_{-}^{b}\tilde{\psi}_{+}^{c}\tilde{\psi}_{-}^{a}
\notag\\ = &-
\frac{i}{2}\tilde{R}^{i}_{mnk}T_{b}^{n}T_{c}^{k}T_{a}^{m}
\psi_{+}^{b} \psi_{+}^{c}\partial_{-}x^{a} +
\frac{i}{2}\tilde{R}^{i}_{mnk}T_{b}^{n}T_{c}^{k}T_{a}^{m}
\psi_{-}^{b} \psi_{-}^{c}\partial_{+}x^{a} \notag\\ &- \frac{1}{2}
\tilde{D}^{i}\tilde{R}_{jkmn}T_{d}^{n}T_{b}^{k}T_{c}^{m}T_{a}^{j}\psi_{+}^{d}\psi_{-}^{b}\psi_{+}^{c}\psi_{-}^{a}
\label{equation4.112}
\end{align}
A quick comparison shows that we obtain equation
(\ref{equation4.108}), and $\tilde{D}^{i}\tilde{R}_{jkmn} = 0$. We
notice that covariant derivatives of curvatures on both spaces
vanish while curvatures are constants, and related to each other by
(\ref{equation4.108}). This obeys that both models are based on
symmetric spaces.

\subsection{Orthonormal Coframes} \label{sec4:OC2}

Equations of motion following from (\ref{equation4.75}) are
\begin{equation}
X_{-+}^{k} = - [\mathbf{\Gamma}_{ij}^{k}(X) -
\mathcal{H}_{ij}^{k}(X)]X_{+}^{i}X_{-}^{j} \label{equation4.113}
\end{equation}
where $X_{+} = D_{+}X$, $X_{-} = D_{-}X$ and $X_{-+} = D_{+}D_{-}X$.
On the contrary to (1, 0) case, this time one writes that $X_{-+} =
-X_{+-}$ and $\{X_{+}, X_{-}\} = 0$, where $\{, \}$ defines the
anticommutation. Superspace coordinates are $z = (\sigma^{\pm},
\theta^{\pm})$, and orthonormal frame can be chosen as $\{
\Lambda^{i} \}$ with connection one form $\{ \Lambda_{j}^{i} \}$.
Similar to (\ref{equation4.46}) and (\ref{equation4.47}) one form
$\{\Lambda^{i}\}$ and covariant derivative of $X_{M}$ can be written
as
\begin{align}
\Lambda^{i} &= dz^{M}X_{M}^{i} \label{equation4.114}\\
dX_{M}^{i} + \Lambda_{j}^{i}X_{M}^{j} &= dz^{N}X_{MN}^{i}
\label{equation4.115}
\end{align}

Pseudoduality relations are
\begin{equation}
\tilde{X}_{\pm}^{i} = \pm \mathcal{T}_{j}^{i} X_{\pm}^{j}
\label{equation4.116}
\end{equation}
We are going to mimic the calculations performed in (1, 0) case
except notable differences $dz^{+} \wedge dz^{-} = - dz^{-} \wedge
dz^{+}$, $X_{+-} = - X_{-+}$, and $X_{+}X_{-} = - X_{-}X_{+}$. We
take exterior derivative of (\ref{equation4.116}), and then use
(\ref{equation4.115}) for both manifolds, and arrange the terms to
get

\begin{equation}
dz^{N}\tilde{X}_{\pm N}^{i} = \pm (d\mathcal{T}_{k}^{i} -
\mathcal{T}_{j}^{i} \Lambda_{k}^{j} + \tilde{\Lambda}_{j}^{i}
\mathcal{T}_{k}^{j})X_{\pm}^{k} \pm dz^{N} \mathcal{T}_{j}^{i}X_{\pm
N}^{j} \label{equation4.117}
\end{equation}
We wedge the plus equation by $dz^{+}$ and minus equation by
$dz^{-}$ to get
\begin{align}
dz^{+} \wedge dz^{-} \tilde{X}_{+-} = dz^{+} \wedge
(d\mathcal{T}_{k}^{i} - \mathcal{T}_{j}^{i}\Lambda_{k}^{j} +
\tilde{\Lambda}_{j}^{i}\mathcal{T}_{k}^{j})X_{+}^{k} + dz^{+} \wedge
dz^{-}
\mathcal{T}_{j}^{i}X_{+-}^{j} \label{equation4.118}\\
dz^{-} \wedge dz^{+} \tilde{X}_{-+} = - dz^{-} \wedge
(d\mathcal{T}_{k}^{i} - \mathcal{T}_{j}^{i}\Lambda_{k}^{j} +
\tilde{\Lambda}_{j}^{i}\mathcal{T}_{k}^{j})X_{-}^{k} - dz^{-} \wedge
dz^{+} \mathcal{T}_{j}^{i}X_{-+}^{j} \label{equation4.119}
\end{align}
we set left-hand sides equal to each other using $\tilde{X}_{+-} = -
\tilde{X}_{-+}$ and $dz^{+} \wedge dz^{-} = - dz^{-} \wedge dz^{+}$.
We notice that we have symmetric expression which has antisymmetric
terms in pairs. Therefore expressions from (\ref{equation4.53}) to
(\ref{equation4.73}) can be repeated. This ends up with the same
result, curvatures of the supersymmetric sigma models will be
constant and opposite to each other, yielding the dual symmetric
spaces.

\section{Pseudoduality in Super WZW Models}\label{sec:int4}

At this point it is interesting to discuss the pseudoduality
transformations on super WZW models \cite{Evans1}. The super WZW
model has considerable interest in the context of conformal field
theory. We use the superspace with coordinates ($\sigma^{+}$,
$\sigma^{-}$, $\theta^{+}$, $\theta^{-}$) where $\sigma^{\pm}$ are
the standard lightcone coordinates, and $\theta^{\pm}$ are the real
Grassmann numbers, with supercharges $Q_{\pm} =
\partial_{\theta^{\pm}} - i\theta^{\pm}\partial_{\pm}$ and
supercovariant derivatives $D_{\pm} = \partial_{\theta^{\pm}} +
i\theta^{\pm}\partial_{\pm}$. To define super WZW model we introduce
the superfield $\mathcal{G} (\sigma, \theta)$ in $G$ with components
as expanded by
\begin{equation}
\mathcal{G }(\sigma, \theta) = g(\sigma)(1 +
i\theta^{+}\psi_{+}(\sigma) + i\theta^{-}\psi_{-}(\sigma) +
i\theta^{+}\theta^{-}\chi(\sigma)) \label{equa4.1}
\end{equation}
where the fermions $\psi_{\pm}(\sigma)$ take values in $\textbf{g}$,
and are the superpartners of the group-valued fields $g(\sigma)$.
The field $\chi(\sigma)$ is the auxiliary field. The lagrangian of
the model can be written as
\begin{equation}
\mathcal{L} = \frac{1}{2}Tr(D_{+}\mathcal{G}^{-1} D_{-}\mathcal{G})
+ \mathbf{\Gamma} \label{equa4.2}
\end{equation}
where $\mathbf{\Gamma}$ represents the WZ term. Equations of motion
following from this lagrangian are
\begin{align}
D_{-}(\mathcal{G}^{-1}D_{+}\mathcal{G}) &= 0 \label{equa4.3}\\
D_{+}[(D_{-}\mathcal{G})\mathcal{G}^{-1}] &= 0 \label{equa4.4}
\end{align}

There is a global symmetry $G_{L} \times G_{R}$ which gives the
conserved super currents $\mathcal{J}_{+}^{L} =
\mathcal{G}^{-1}D_{+}\mathcal{G}$ and $\mathcal{J}_{-}^{R} =
(D_{-}\mathcal{G})\mathcal{G}^{-1}$.

We can write similar expressions related to pseudodual WZW model
with tilde. One can write the pseudoduality transformations using
the similarity with bosonic case
\begin{align}
\tilde{\mathcal{G}}^{-1}D_{+}\tilde{\mathcal{G}} &= +\mathcal{T}(\sigma, \theta) \mathcal{G}^{-1}D_{+}\mathcal{G} \label{equa4.5}\\
\tilde{\mathcal{G}}^{-1}D_{-}\tilde{\mathcal{G}} &=
-\mathcal{T}(\sigma, \theta) \mathcal{G}^{-1}D_{-}\mathcal{G}
\label{equa4.6}
\end{align}

Taking $D_{-}$ of first equation (\ref{equa4.5}) followed by
(\ref{equa4.3}) yields that $D_{-}\mathcal{T}(\sigma, \theta) = 0$.
If $\mathcal{T}(\sigma, \theta)$ is expanded as $\mathcal{T}
(\sigma, \theta) = T(\sigma) + \theta^{+}\lambda_{+} +
\theta^{-}\lambda_{-} + \theta^{+}\theta^{-}N(\sigma)$, then the
condition $D_{-}\mathcal{T}(\sigma, \theta)$ implies that
$\lambda_{-} = 0$, $N(\sigma) = 0$, $\partial_{-}T(\sigma) = 0$ and
$\partial_{-}\lambda_{+} = 0$. Hence $\mathcal{T}$ turns out to be
\begin{equation}
\mathcal{T}(\sigma, \theta) = T(\sigma^{+}) +
\theta^{+}\lambda_{+}(\sigma^{+}) \label{equa4.7}
\end{equation}

Taking $D_{+}$ of second equation (\ref{equa4.6}) gives the
following equation
\begin{equation}
D_{+}\mathcal{T}_{j}^{j} (\sigma, \theta) =
(\tilde{f}_{mn}^{i}\mathcal{T}_{j}^{m}\mathcal{T}_{k}^{n} -
f_{jk}^{m}\mathcal{T}_{m}^{i})(\mathcal{G}^{-1}D_{+}\mathcal{G})^{k}
\label{equa4.8}
\end{equation}
Before going further to solve this equation, it is convenient to
find out the values of some fields in terms of components. A brief
computation shows that
\begin{align}
\mathcal{G}^{-1}D_{+}\mathcal{G} = &i\psi_{+} + i\theta^{+}
(g^{-1}\partial_{+}g -i
\psi_{+}^{2}) + i \theta^{-}(\chi -i\psi_{-}\psi_{+}) \label{equa4.9}\\
&- \theta^{+}\theta^{-}(\partial_{+}\psi_{-} +
[g^{-1}\partial_{+}g, \psi_{-}] + [\psi_{+}, \chi]) \notag\\
\mathcal{G}^{-}D_{-}\mathcal{G} = &i\psi_{-} - i\theta^{+}(\chi
-i\psi_{-}\psi_{+}) +
i\theta^{-} (g^{-1}\partial_{-}g -i\psi_{-}^{2}) \label{equa4.10}\\
&+ \theta^{+}\theta^{-}(\partial_{-}\psi_{+} +
[g^{-1}\partial_{-}g, \psi_{+}] + [\chi, \psi_{-}]) \notag\\
(D_{-}\mathcal{G})\mathcal{G}^{-1} = &g\{i\psi_{-} -
i\theta^{+}(\chi -i\psi_{-}\psi_{+})
+ i\theta^{-} (g^{-1}\partial_{-}g + i\psi_{-}^{2}) \label{equa4.11}\\
&+ \theta^{+}\theta^{-} (\partial_{-}\psi_{+} + [\psi_{-},
\chi]\}g^{-1} \notag
\end{align}

Hence, the equation of motion (\ref{equa4.3}) produces the following
equations
\begin{align}
\chi &= i\psi_{-}\psi_{+} \label{equa4.12}\\
\partial_{-}\psi_{+} &= 0 \label{equa4.13}\\
\partial_{-} (g^{-1}\partial_{+}g -i \psi_{+}^{2}) &= 0
\label{equa4.14}\\
\partial_{+}\psi_{-} &= [\psi_{-}, g^{-1}\partial_{+}g] + [\chi, \psi_{+}]
\label{equa4.15}
\end{align}
and (\ref{equa4.4}) yields that
\begin{align}
\chi &= i\psi_{-}\psi_{+} \label{equa4.16}\\
\partial_{-}\psi_{+} &= [\chi, \psi_{-}] \label{equa4.17}\\
\partial_{+}\psi_{-} &= [\psi_{-}, g^{-1}\partial_{+}g]
\label{equa4.18}\\
\partial_{+}(g^{-1}\partial_{-}g + i\psi_{-}^{2}) &= [g^{-1}\partial_{-}g + i\psi_{-}^{2},
g^{-1}\partial_{+}g] \label{equa4.19}
\end{align}

We see that (\ref{equa4.12}) and (\ref{equa4.16}) are the same
expressions, and determines the auxiliary field in terms of
$\psi_{-}$ and $\psi_{+}$. (\ref{equa4.13}) implies that $\psi_{+}$
depends on $\sigma^{+}$ only, and (\ref{equa4.17}) points out that
$\chi$ commutes with $\psi_{-}$ as expected. (\ref{equa4.14}) gives
us the bosonic left current conservation law by means of
(\ref{equa4.13}). Comparison of (\ref{equa4.15}) with
(\ref{equa4.18}) shows that $\chi$ commutes with $\psi_{+}$, and
(\ref{equa4.18}) is the fermionic equation of motion for $\psi_{-}$,
 which leads (\ref{equa4.19}) to the bosonic right current conservation
 law. Finally we may eliminate $\psi_{\pm}^{2}$ terms because these
 are fermionic fields and anticommute with each other.

Therefore the fields (\ref{equa4.9})-(\ref{equa4.11}) can be written
in simplified forms as
\begin{align}
\mathcal{G}^{-1}D_{+}\mathcal{G} &= i\psi_{+} + i \theta^{+} g^{-1}\partial_{+}g \label{equa4.20}\\
\mathcal{G}^{-1}D_{-}\mathcal{G} &= i\psi_{-} + i \theta^{-}
g^{-1}\partial_{-}g  + \theta^{+}\theta^{-} [g^{-1}\partial_{-}g,
\psi_{+}]
\label{equa4.21}\\
(D_{-}\mathcal{G})\mathcal{G}^{-1} &= ig\psi_{-}g^{-1} + i
\theta^{-} (\partial_{-}g) g^{-1} \label{equa4.22}
\end{align}

We can now solve the equation (\ref{equa4.8}) using (\ref{equa4.7})
and (\ref{equa4.20}). A little computation gives the components of
$\mathcal{T}(\sigma, \theta)$ as
\begin{align}
(\lambda_{+})_{j}^{i} = &i(\tilde{f}_{mn}^{i}T_{j}^{m}T_{k}^{n} -
f_{jk}^{m}T_{m}^{i}) \psi_{+}^{k} \label{equa4.23}\\
(\partial_{+}T)_{j}^{i} =
&(\tilde{f}_{mn}^{i}(\lambda_{+})_{j}^{m}T_{k}^{n} +
\tilde{f}_{mn}^{i}T_{j}^{m}(\lambda_{+})_{k}^{n} -
f_{jk}^{m}(\lambda_{+})_{m}^{i}) \psi_{+}^{k}\label{equa4.24}\\ &+
(\tilde{f}_{mn}^{i}T_{j}^{m}T_{k}^{n} -
f_{jk}^{m}T_{m}^{i})(g^{-1}\partial_{+}g)^{k} \notag
\end{align}
If (\ref{equa4.23}) is inserted in (\ref{equa4.24}) the result
follows
\begin{align}
(\partial_{+}T)_{j}^{i} = &(\tilde{f}_{mn}^{i}T_{j}^{m}T_{k}^{n} -
f_{jk}^{m}T_{m}^{i})(g^{-1}\partial_{+}g)^{k} - i \tilde{f}_{kl}^{i}
\tilde{f}_{mn}^{l} (T_{j}^{m}T_{b}^{k} -
T_{j}^{k}T_{b}^{m})T_{a}^{n}\psi_{+}^{a}\psi_{+}^{b}\notag\\ &- i
\tilde{f}_{kl}^{i} f_{ba}^{m}T_{j}^{k}T_{m}^{l}
\psi_{+}^{a}\psi_{+}^{b} + if_{jb}^{m}f_{ma}^{n}T_{n}^{i}
\psi_{+}^{a} \psi_{+}^{b} \label{equa4.25}
\end{align}
We want to find perturbation solution, and we notice that the order
of the term $g^{-1}\partial_{+}g$ is proportional to the order of
the term $\psi \psi$. We find the following perturbative result up
to the second order terms after integrating (\ref{equa4.25})
\begin{align}
T_{j}^{i} (\sigma^{+}) = T_{j}^{i} (0) +  A_{jk}^{i}
\int_{0}^{\sigma^{+}} (g^{-1}\partial_{+}g)^{k} d\sigma'^{+} +
B_{jab}^{i} \int_{0}^{\sigma^{+}} \psi_{+}^{a} \psi_{+}^{b}
d\sigma'^{+} + H.O. \label{equa4.26}
\end{align}
where $T_{j}^{i} (0) = \delta_{j}^{i}$, $A_{jk}^{i} =
(\tilde{f}_{jk}^{i} - f_{jk}^{i})$, and $B_{jab}^{i} =
i(\tilde{f}_{ak}^{i}\tilde{f}_{bj}^{k} +
\tilde{f}_{jk}^{i}f_{ab}^{k} + f_{ak}^{i}f_{bj}^{k})$. Therefore
$\lambda_{+}$ may be written as
\begin{align}
(\lambda_{+})_{j}^{i} = & i A_{jk}^{i} \psi_{+}^{k} + C_{jkc}^{i}
\psi_{+}^{k} \int_{0}^{\sigma^{+}} (g^{-1}\partial_{+}g)^{c}
d\sigma'^{+}\notag\\ &+ i D_{jkcd}^{i} \psi_{+}^{k}
\int_{0}^{\sigma^{+}} \psi_{+}^{c}\psi_{+}^{d} d\sigma'^{+} + H.O.
\label{equa4.27}
\end{align}
where constants $C_{jkc}^{i}$ and $D_{jkcd}^{i}$ are
\begin{align}
C_{jkc}^{i} = &\tilde{f}_{jn}^{i}A_{kc}^{n} +
\tilde{f}_{nk}^{i}A_{jc}^{n} - f_{jk}^{n}A_{nc}^{i} =
(\tilde{f}_{nc}^{i}\tilde{f}_{jk}^{n} -
\tilde{f}_{n[j}^{i}f_{ck]}^{n} + f_{nc}^{i}f_{jk}^{n})
\label{equa4.28}\\
D_{jkcd}^{i} = &\tilde{f}_{jn}^{i}B_{kcd}^{n} +
\tilde{f}_{nk}^{i}B_{jcd}^{n} - f_{jk}^{n}B_{ncd}^{i} =
i\tilde{f}_{jn}^{i}\tilde{f}_{cm}^{n}\tilde{f}_{kd}^{m} +
i\tilde{f}_{nk}^{i}\tilde{f}_{cm}^{n}\tilde{f}_{jd}^{m} \notag\\ &+
i\tilde{f}_{jn}^{i}\tilde{f}_{km}^{n}f_{da}^{m} +
i\tilde{f}_{nk}^{i}\tilde{f}_{jm}^{n}f_{dc}^{m} -
i\tilde{f}_{cm}^{i}\tilde{f}_{nd}^{m}f_{jk}^{n} +
i\tilde{f}_{jn}^{i}f_{cm}^{n}f_{kd}^{m} \notag\\ &+
i\tilde{f}_{nk}^{i}f_{cm}^{n}f_{jd}^{m} -
i\tilde{f}_{nm}^{i}f_{dc}^{m}f_{jk}^{n}
-if_{cm}^{i}f_{nd}^{m}f_{jk}^{n} \label{equa4.29}
\end{align}
As seen we have an expression for the transformation matrix
(\ref{equa4.7}) up to the third order terms. We notice that $T$
represents even order terms while $\lambda_{+}$ represents odd order
terms. Now we can proceed to find expressions on
$\tilde{\mathbb{M}}$ using pseudoduality equations (\ref{equa4.5})
and (\ref{equa4.6}). If (\ref{equa4.7}) and (\ref{equa4.20}) are
substituted in the first equation we obtain
\begin{align}
\tilde{\psi}_{+}^{i} &= T_{j}^{i}
\psi_{+}^{j} \label{equa4.30}\\
(\tilde{g}^{-1}\partial_{+}\tilde{g})^{i} &=
T_{j}^{i}(g^{-1}\partial_{+}g)^{j}
+(\lambda_{+})_{j}^{i}\psi_{+}^{j} \label{equa4.31}
\end{align}
We notice that both of these equations depend only on $\sigma^{+}$.
Likewise inserting (\ref{equa4.7}) and (\ref{equa4.21}) into second
equation (\ref{equa4.6}) leads to
\begin{align}
(\lambda_{+})_{j}^{i} \psi_{-}^{j} &= 0 \label{equa4.32}\\
\tilde{\psi}_{-}^{i} &= - T_{j}^{i}\psi_{-}^{j} \label{equa4.33}\\
(\tilde{g}^{-1}\partial_{-}\tilde{g})^{i} &= -
T_{j}^{i}(g^{-1}\partial_{-}g)^{j} \label{equa4.34}\\
[\tilde{g}^{-1}\partial_{-}\tilde{g}, \tilde{\psi}_{+}]^{i} &= -
T_{j}^{i} [g^{-1}\partial_{-}g, \psi_{+}]^{j} + i
(\lambda_{+})_{j}^{i}(g^{-1}\partial_{-}g)^{j} \label{equa4.35}
\end{align}
These are the pseudoduality equations in components. We observe that
if $\psi_{-}$ and $\psi_{+}$ are set to zero we obtain bosonic case
pseudoduality equations as pointed out in (\cite{alvarez2}). We see
that the term $(\lambda_{+})_{j}^{i}\psi_{+}^{j}$ in equation
(\ref{equa4.31}) gives us $(\lambda_{+})_{j}^{i}\psi_{+}^{j} = -i
[\tilde{\psi}_{+}, \tilde{\psi}_{+}]_{\tilde{G}}^{i} + i
T_{j}^{i}[\psi_{+}, \psi_{+}]_{G}^{j} = 0$. The last equation
(\ref{equa4.35}) gives us the constraint (\ref{equa4.23}). The
equation (\ref{equa4.32}) is interesting because it tells us that
$[\tilde{\psi}_{-}, \tilde{\psi}_{+}]^{i} = - T_{j}^{i} [\psi_{-},
\psi_{+}]^{j}$, which gives us two choices. First choice is
$\lambda_{+} = 0$ which leads to either
\begin{equation}
\tilde{f}_{mn}^{i} T_{k}^{m} T_{l}^{n} = T_{j}^{i} f_{kl}^{j}
\label{equa4.36}
\end{equation}
if $\psi_{+} \neq 0$. This yields that $\partial_{+}T = 0$ as can be
seen from (\ref{equa4.24}), and hence we get a trivial case, flat
space pseudoduality equations as follows
\begin{align}
\tilde{\psi}_{\pm}^{i} &= \pm \psi_{\pm}^{i} \label{equa4.37}\\
(\tilde{g}^{-1}\partial_{\pm}\tilde{g})^{i} &= \pm
(g^{-1}\partial_{\pm}g)^{i} \label{equa4.38}
\end{align}
where we choose $T$ to be identity. Therefore we obtain
$\tilde{f}_{jk}^{i} = f_{jk}^{i}$ in (\ref{equa4.36}). Or we set
$\psi_{+} = 0$, and hence last term in (\ref{equa4.26}) will be
eliminated, so pseudoduality relations will be
\begin{align}
\tilde{\psi}_{-}^{i} = &- \psi_{-}^{i} - [\psi_{-},
\int_{0}^{\sigma^{+}}
(g^{-1}\partial_{+}g)d\sigma'^{+}]_{\tilde{G}}^{i}
\label{equa4.39}\\ &+ [\psi_{-}, \int_{0}^{\sigma^{+}}
(g^{-1}\partial_{+}g)d\sigma'^{+}]_{G}^{i} + H.O. \notag
\end{align}
\begin{align}
(\tilde{g}^{-1}\partial_{\pm}\tilde{g})^{i} = &\pm
(g^{-1}\partial_{\pm}g)^{i} \pm [g^{-1}\partial_{\pm}g,
\int_{0}^{\sigma^{+}} (g^{-1}\partial_{+}g)
d\sigma'^{+}]_{\tilde{G}}^{i} \label{equa4.40}\\ &\mp
[g^{-1}\partial_{\pm}g, \int_{0}^{\sigma^{+}} (g^{-1}\partial_{+}g)
d\sigma'^{+}]_{G}^{i} + H.O. \notag
\end{align}
where we introduced the bracket $[\ ,\ ]_{G/\tilde{G}}$ to represent
the commutations in $G/\tilde{G}$. Second choice will eliminate
$\psi_{-}$ and hence we get whole expressions (\ref{equa4.26}) and
(\ref{equa4.27}) for $T$ and $\lambda_{+}$. Therefore we obtain the
following perturbation fields
\begin{align}
\tilde{\psi}_{+}^{i} = & \psi_{+}^{i} + [\psi_{+},
\int_{0}^{\sigma^{+}} (g^{-1}\partial_{+}g)
d\sigma'^{+}]_{\tilde{G}}^{i} - [\psi_{+}, \int_{0}^{\sigma^{+}}
(g^{-1}\partial_{+}g) d\sigma'^{+}]_{G}^{i} \notag\\ &+ i
\int_{0}^{\sigma^{+}} [\psi_{+}(\sigma^{'^{+}}),
[\psi_{+}(\sigma'^{+}),
\psi_{+}(\sigma^{+})]_{\tilde{G}}]_{\tilde{G}}^{i} d\sigma'^{+}
\notag\\ &+ i \int_{0}^{\sigma^{+}} [\psi_{+}(\sigma^{'^{+}}),
[\psi_{+}(\sigma'^{+}), \psi_{+}(\sigma^{+})]_{G}]_{G}^{i}
d\sigma'^{+} + H.O. \label{equa4.41}
\end{align}
\begin{align}
(\tilde{g}^{-1}\partial_{\pm}\tilde{g})^{i} = &\pm
(g^{-1}\partial_{\pm}g)^{i} \pm [g^{-1}\partial_{\pm}g,
\int_{0}^{\sigma^{+}} (g^{-1}\partial_{+}g)
d\sigma'^{+}]_{\tilde{G}}^{i} \mp [g^{-1}\partial_{\pm}g,
\int_{0}^{\sigma^{+}} (g^{-1}\partial_{+}g)
d\sigma'^{+}]_{G}^{i}\notag\\ &\pm i \int_{0}^{\sigma^{+}}
[\psi_{+}(\sigma'^{+}), [\psi_{+}(\sigma'^{+}),
(g^{-1}\partial_{\pm}g)(\sigma^{+})]_{\tilde{G}}]_{\tilde{G}}^{i}
d\sigma'^{+} \notag\\ &\pm i \int_{0}^{\sigma^{+}}
[\psi_{+}(\sigma'^{+}), [\psi_{+}(\sigma'^{+}),
(g^{-1}\partial_{\pm}g)(\sigma^{+})]_{G}]_{G}^{i} d\sigma'^{+} +
H.O. \label{equa4.42}
\end{align}
where the cross terms $[\ , [\ ,\ ]_{G}]_{\tilde{G}}$ vanish.

 We have already derived our pseudoduality equations,
conditions inducing pseudoduality, and finally the perturbative
expressions of the pseudodual fields up to the third (fourth) order
terms, leading to conserved currents on the pseudodual model. Using
these fields it is possible to construct left and right super
currents on pseudodual manifold $\tilde{G}$. It is apparent from the
expression (\ref{equa4.20}) that we can easily construct right super
currents belonging to special cases discussed above. To find left
super currents we use the method we traced in \cite{msarisaman1,
msarisaman2}.

\subsection{Supercurrents in Flat Space Pseudoduality}
\label{SCFSP}

In this case structure constants of both models are the same,
$\tilde{f} = f$, and pseudoduality relations are given by
(\ref{equa4.37}) and (\ref{equa4.38}). We let $g = e^{Y}$, where $Y$
is the lie algebra. Using the expansion \cite{helgason, forger1,
forger3, msarisaman2}
\begin{equation}
g^{-1}\partial_{\pm}g = \frac{1 - e^{-adY}}{adY} \partial_{\pm} =
\sum_{k = 0}^{\infty} \frac{(-1)^{k}}{(k + 1)!} [Y, ...,[Y,
\partial_{\pm}Y]] \label{equa4.43}
\end{equation}
where $adY$ is the adjoint representation of $Y$, and $ad Y (Z) =
[Y, Z]$. We know that bosonic currents are invariant under $g
\longrightarrow g_{R} (\sigma^{-}) g_{L} (\sigma^{+})$, hence we
obtain that $g^{-1}\partial_{+}g \longrightarrow
g_{L}^{-1}\partial_{+}g_{L}$, which is
\begin{equation}
g_{L}^{-1}\partial_{+}g_{L} = \partial_{+}Y_{L} - \frac{1}{2!}
[Y_{L}, \partial_{+}Y_{L}] + \frac{1}{3!} [Y_{L}, [Y_{L},
\partial_{+}Y_{L}]] + ...
\end{equation}

Now we impose that $Y = \sum_{0}^{\infty} \varepsilon^{n}y_{n}$,
where $\varepsilon$ is a small parameter. Thus we get the following
lie algebra valued field up to the third order terms
\begin{align}
g_{L}^{-1}\partial_{+}g_{L} = &\varepsilon \partial_{+}y_{L1} +
\varepsilon^{2} (\partial_{+}y_{L2} - \frac{1}{2}[y_{L1},
\partial_{+}y_{L1}]) \label{equa4.45}\\ &+ \varepsilon^{3} (\partial_{+}y_{L3} - \frac{1}{2}[y_{L1}, \partial_{+}y_{L2}] - \frac{1}{2}[y_{L2}, \partial_{L1}] + \frac{1}{6} [y_{L1}, [y_{L1},
\partial_{+}y_{L1}]]) + \mathcal{O} (\varepsilon^{4}) \notag
\end{align}
In a similar way one can find the  expression for
$g^{-1}\partial_{-}g$ \cite{msarisaman2}
\begin{align}
g^{-1}\partial_{-}g = &\varepsilon \partial_{-}y_{R1} +
\varepsilon^{2} (\partial_{-}y_{R2} - [y_{L1}, \partial_{-}y_{R1}] -
\frac{1}{2} [y_{R1}, \partial_{-}y_{R1}]) \label{equa4.46}\\ &+
\varepsilon^{3} (\partial_{-}y_{R3} - [y_{L2}, \partial_{-}y_{R1}] -
[y_{L1}, \partial_{-}y_{R2}] - \frac{1}{2} [y_{R2},
\partial_{-}y_{R1}] - \frac{1}{2} [y_{R1}, \partial_{-}y_{R2}])
\notag\\ &+ \frac{1}{2} [y_{L1}, [y_{R1}, \partial_{-}y_{R1}]] +
\frac{1}{2} [y_{L1}, [y_{L1}, \partial_{-}y_{R1}]] + \mathcal{O}
(\varepsilon^{4}) \notag
\end{align}

Since it works all the way up we are going to do all our
calculations up to the second order of $\varepsilon$ for simplicity
and demonstration. We can write similar expressions for the manifold
$\tilde{G}$. Pseudoduality equation (\ref{equa4.38}) gives infinite
number of sub-pseudoduality equations, from which we may write the
following expressions coming from up to the second order of
$\varepsilon$ terms
\begin{align}
\partial_{+}\tilde{y}_{L1} &= \partial_{+}y_{L1} \label{equa4.47}\\
\partial_{-}\tilde{y}_{R1} &= - \partial_{-}y_{R1} \label{equa4.48}\\
\partial_{+}\tilde{y}_{L2} - \frac{1}{2}[\tilde{y}_{L1},
\partial_{+}\tilde{y}_{L1}] &= \partial_{+}y_{L2} - \frac{1}{2}[y_{L1},
\partial_{+}y_{L1}] \label{equa4.49}\\
\partial_{-}\tilde{y}_{R2} - [\tilde{y}_{L1}, \partial_{-}\tilde{y}_{R1}] -
\frac{1}{2} [\tilde{y}_{R1}, \partial_{-}\tilde{y}_{R1}] &= -
\partial_{-}y_{R2} + [y_{L1}, \partial_{-}y_{R1}] + \frac{1}{2}
[y_{R1},
\partial_{-}y_{R1}] \label{equa4.50}
\end{align}
First equation yields that $\tilde{y}_{L1} = y_{L1} + C_{L1}$, where
$C_{L1}$ is constant, and the second equation gives $\tilde{y}_{R1}
= - y_{R1} - C_{R1}$, where $C_{R1}$ is constant. Inserting these
result into last equation gives
\begin{equation}
\partial_{-}\tilde{y}_{R2} + \frac{1}{2} [\tilde{y}_{R1}, \partial_{-}\tilde{y}_{R1}] = -
\partial_{-}y_{R2} + \frac{3}{2} [y_{R1}, \partial_{-}y_{R1}]
\label{equa4.51}
\end{equation}
where we used the equality of structure constants. We found this
because we need this term in the expansion of bosonic right current
\footnote{see (\cite{msarisaman2}) \cite{helgason, forger1, forger3}
for details of this expansion}, which is
\begin{equation}
(\partial_{-}g_{R})g_{R}^{-1} = \varepsilon \partial_{-} y_{R1} +
\varepsilon^{2} (\partial_{-}y_{R2} + \frac{1}{2} [y_{R1},
\partial_{-}y_{R1}]) + \mathcal{O} (\varepsilon^{3}) \label{equa4.52}
\end{equation}
Hence bosonic right and left currents on $\tilde{G}$ in terms of
nonlocal expressions will be
\begin{align}
\tilde{J}_{+}^{L} &= \tilde{g}_{L}^{-1}\partial_{+}\tilde{g}_{L} =
\varepsilon \partial_{+}y_{L1} + \varepsilon^{2} (\partial_{+}y_{L2}
- \frac{1}{2}[y_{L1},
\partial_{+}y_{L1}]) + \mathcal{O} (\varepsilon^{3})
\label{equa4.53}\\
\tilde{J}_{-}^{R} &= (\partial_{-}\tilde{g}_{R})\tilde{g}_{R}^{-1} =
- \varepsilon \partial_{-}y_{R1} - \varepsilon^{2}
(\partial_{-}y_{R2} - \frac{3}{2} [y_{R1}, \partial_{-}y_{R1}]) +
\mathcal{O} (\varepsilon^{3}) \label{equa4.54}
\end{align}
Obviously these currents are conserved by means of (\ref{equa4.14})
and (\ref{equa4.19}). Now we consider the fermionic components, and
we let $\psi_{\pm} = \sum_{n = 1}^{\infty} \varepsilon^{n} \psi_{n
\pm}$. We denote $\psi_{\pm}$ as the sum of right and left
components $\psi_{\pm} = \psi_{R \pm} (\sigma^{-}) + \psi_{L \pm}
(\sigma^{+})$. But from (\ref{equa4.13}) we understand that
$\psi_{+}$ includes $\psi_{L +}$ only. Pseudoduality relations
(\ref{equa4.37}) again yields infinite number of subequations
\begin{align}
\tilde{\psi}_{L n +} &= \psi_{L n +} \label{equa4.55}\\
\tilde{\psi}_{(L/R)n -} &= - \psi_{(L/R)n -} \label{equa4.56}
\end{align}
which hold true for each $n$. Thus left and right supercurrents on
$\tilde{G}$ in nonlocal terms up to the second order of
$\varepsilon$ will be
\begin{align}
\mathcal{\tilde{J}}_{+}^{L} &=
\tilde{\mathcal{G}}^{-1}D_{+}\tilde{\mathcal{G}} = i
\tilde{\psi}_{+} +
i\theta^{+} (\tilde{g}^{-1}\partial_{+}\tilde{g}) \label{equa4.57}\\
&= i\varepsilon (\psi_{L 1 +} + \theta^{+}
\partial_{+}y_{L1}) + i \varepsilon^{2} \{\psi_{L 2 +} + \theta^{+} (\partial_{+}y_{L2} - \frac{1}{2}[y_{L1},
\partial_{+}y_{L1}])\} + \mathcal{O} (\varepsilon^{3}) \notag
\end{align}
\begin{align}
\mathcal{\tilde{J}}_{-}^{R} &=
(D_{-}\tilde{\mathcal{G}})\tilde{\mathcal{G}}^{-1} = i\tilde{g}
\tilde{\psi}_{-} \tilde{g}^{-1} + i\theta^{-}
(\partial_{-}\tilde{g})\tilde{g}^{-1} \label{equa4.58}\\
& = - i\varepsilon (\psi_{1 -} + \theta^{-} \partial_{-}y_{R1}) - i
\varepsilon^{2} \{\psi_{2 -} + [y_{L1}, \psi_{1 -}] - [y_{R1},
\psi_{1 -}] \notag\\ &+ \theta^{-} (\partial_{-}y_{R2} - \frac{3}{2}
[y_{R1},
\partial_{-}y_{R1}])\} + \mathcal{O} (\varepsilon^{3}) \notag
\end{align}
It is obvious from the equations of motion that these currents in
nonlocal expressions are conserved.

\subsection{Supercurrents in Anti-chiral Pseudoduality}
\label{SACP}

Now we consider our second case where $\psi_{+}$ vanishes. In this
case we need to be careful when using bracket relations because
structure constants are different. We have already found our
nonlocal expressions in (\ref{equa4.39}) and (\ref{equa4.40}). We
use the same expansions of lie algebra $Y$ and fermionic field
$\psi_{-}$ in the powers of $\varepsilon$ as used in the previous
part. Therefore pseudoduality relations up to the second order of
$\varepsilon$ yield the following equations
\begin{align}
\tilde{\psi}_{1 -}^{i} &= - \psi_{1 -}^{i} \label{equa4.59}\\
\tilde{\psi}_{2 -}^{i} &= - \psi_{2 -}^{i} - [\psi_{1 -},
y_{L1}]_{\tilde{G}}^{i} + [\psi_{1 -}, y_{L1}]_{G}^{i}
\label{equa4.60}
\end{align}
\begin{align}
\partial_{+}\tilde{y}_{L1}^{i} &= \partial_{+}y_{L1}^{i}
\label{equa4.61}\\
\partial_{+}\tilde{y}_{L2}^{i} - [\tilde{y}_{L1},
\partial_{+}\tilde{y}_{L1}]_{\tilde{G}}^{i} &= \partial_{+}y_{L2}^{i} +
\frac{1}{2} [y_{L1}, \partial_{+}y_{L1}]_{G}^{i} - [y_{L1},
\partial_{+}y_{L1}]_{\tilde{G}}^{i} \label{equa4.62}
\end{align}
\begin{align}
\partial_{-}\tilde{y}_{R1}^{i} &= - \partial_{-}y_{R1}^{i} \label{equa4.63}\\
\partial_{-}\tilde{y}_{R2}^{i} + \frac{1}{2} [\tilde{y}_{R1},
\partial_{-}\tilde{y}_{R1}]_{\tilde{G}}^{i} &= - \partial_{-}y_{R2}^{i} +
\frac{1}{2} [y_{R1}, \partial_{-}y_{R1}]_{G}^{i} + [y_{R1},
\partial_{-}y_{R1}]_{\tilde{G}}^{i} \label{equa4.64}
\end{align}
We may find out nonlocal supercurrents on the pseudodual manifold
using these expressions
\begin{align}
\mathcal{\tilde{J}}_{+}^{L} &= i \varepsilon \theta^{+}
\partial_{+}y_{L1} + i \varepsilon^{2} \theta^{+} \{\partial_{+}y_{L2} + \frac{1}{2} [y_{L1}, \partial_{+}y_{L1}]_{G}^{i} - [y_{L1},
\partial_{+}y_{L1}]_{\tilde{G}}^{i}\} + \mathcal{O}
(\varepsilon^{3}) \label{equa4.65}\\
\mathcal{\tilde{J}}_{-}^{R} &= - i \varepsilon (\psi_{1 -} +
\theta^{-} \partial_{-}y_{R1}) - i \varepsilon^{2} \{\psi_{2 -} +
[y_{L1}, \psi_{1 -}]_{G} - [y_{R1}, \psi_{1 -}]_{\tilde{G}}
\label{equa4.66}\\ &+ \theta^{-} (\partial_{-}y_{R2} - \frac{1}{2}
[y_{R1},
\partial_{-}y_{R1}]_{G} - [y_{R1}, \partial_{-}y_{R1}]_{\tilde{G}})
\} + \mathcal{O} (\varepsilon^{3}) \notag
\end{align}
Obviously these currents in nonlocal expressions are conserved
provided that equations of motion are satisfied.

\subsection{Supercurrents in Chiral Pseudoduality} \label{SACP}

We consider our final case where $\psi_{-}$ disappears. We notice
that there is a contribution of chiral part in the isometry $T$
which leads to third order terms in the field expressions on the
target space of pseudodual manifold as can be seen from equations
(\ref{equa4.41}) and (\ref{equa4.42}). Again we keep in our minds
that structure constants are different. If the same conventions for
$Y$ and $\psi_{+}$ are used as above, then pseudoduality relations
up to the second order of $\varepsilon$ can be calculated.
Expressions for the fields $\tilde{g}^{-1}\partial_{\pm}\tilde{g}$
are the same as (\ref{equa4.61})-(\ref{equa4.64}), and expression
for the chiral field (\ref{equa4.41}) gives that
\begin{align}
\tilde{\psi}_{L 1 +}^{i} &= \psi_{L 1 +}^{i} \label{equa4.67}\\
\tilde{\psi}_{L 2 +}^{i} &= \psi_{L 2 +}^{i} + [\psi_{L 1 +},
y_{L1}]_{\tilde{G}}^{i} - [\psi_{L 1 +}, y_{L1}]_{G}^{i}
\label{equa4.68}
\end{align}

Then nonlocal conserved supercurrents are found to be
\begin{align}
\mathcal{\tilde{J}}_{+}^{L} &= i \varepsilon (\psi_{L 1 +} +
\theta^{+}\partial_{+}y_{L1}) + i \varepsilon^{2} \{\psi_{L 2 +} +
\theta^{+} (\partial_{+}y_{L2} + \frac{1}{2} [y_{L1},
\partial_{+}y_{L2}]_{G} \label{equa4.69}\\ &- [y_{L1},
\partial_{+}y_{L1}]_{\tilde{G}})\} + \mathcal{O} (\varepsilon^{3}) \notag\\
\mathcal{\tilde{J}}_{-}^{R} &= - i \varepsilon \theta^{-}
\partial_{-}y_{R1} - i \varepsilon^{2} \theta^{-} (\partial_{-}y_{R2} - \frac{1}{2}[y_{R1}, \partial_{-}y_{R1}]_{G} - [y_{R1},
\partial_{-}y_{R1}]_{\tilde{G}}) + \mathcal{O} (\varepsilon^{3})
\label{equa4.70}
\end{align}

It is noted that all these supercurrents are the complements of each
other, and special cases of a more general one. Under the limiting
conditions they are equal to each other. If we denote the bosonic
and fermionic components by $\tilde{J}_{B}$ and $\tilde{J}_{F}$ then
they are written as
\begin{align}
\mathcal{\tilde{J}}_{\pm}^{L/R} &= \pm \tilde{J}_{F}^{L/R} \pm
\theta^{\pm} \tilde{J}_{B}^{L/R} \label{equa4.71}
\end{align}

Since these super currents serve as the orthonormal frame on the
pullback bundle of the target space of $G$, we may find the
corresponding bosonic and fermionic curvatures using them. If $L^{i}
= \mathcal{J}^{i}$ is the left invariant Cartan one form which
satisfies the Maurer-Cartan equation
\begin{equation}
d\mathcal{J}^{i} + \frac{1}{2} f_{jk}^{i} \mathcal{J}^{j} \wedge
\mathcal{J}^{k} = 0\label{equa4.72}
\end{equation}
and $L_{k}^{i} = \frac{1}{2} f_{jk}^{i} \mathcal{J}^{j}$ is the
antisymmetric riemannian connection, then Cartan structural
equations on superspace can be written as
\begin{align}
dL^{i} + L_{j}^{i} \wedge L^{j} &= 0 \label{equa4.73}\\
dL_{j}^{i} + L_{k}^{i} \wedge L_{j}^{k} &= \frac{1}{2}
\mathcal{R}_{jkl}^{i} L^{k} \wedge L^{l} \label{equa4.74}
\end{align}
where $\mathcal{R}_{jkl}^{i}$ is the curvature of superspace. If the
calculations in the previous section is repeated using these
equations in this case one can show that curvatures on $SO(G)$ and
$SO(\tilde{G})$ are constants, and related to each other by
$\tilde{\mathcal{R}}_{jkl}^{i} = - \mathcal{R}_{jkl}^{i}$, which
shows that two superspaces are dual symmetric spaces. If this
curvature relation is split into bosonic and fermionic parts, it is
easy to see that fermionic part will yield a curvature relation
which are opposite to each other, i.e.
$(\tilde{\mathcal{R}}_{F})_{jkl}^{i} = -
(\mathcal{R}_{F})_{jkl}^{i}$, while bosonic part will give that both
curvatures will be the same, i.e.
$(\tilde{\mathcal{R}}_{B})_{jkl}^{i} = (\mathcal{R}_{B})_{jkl}^{i}$,
because of anticommuting numbers. This is consistent with the
results found in the component expansion methods.

\section{Discussion} \label{discussion}

We analyzed the pseudoduality conditions on the supersymmetric
extensions of sigma models in two respects, by component expansion
and orthonormal coframe method. In the first case we have seen that
pseudoduality transformation in N = 1 supersymmetric sigma models
imposes the condition that pseudoduality maps all points in the
first manifold to only one point at which riemann normal coordinates
are used on the pseudodual manifold. Although torsions of both
models vanish in (1, 1) case, torsion of the pseudodual manifold
exists in (1, 0) case. Curvatures when splitted to bosonic and
fermionic parts yield that bosonic curvatures must be the same
because of anticommuting grassmann numbers. It is obvious that
pseudoduality transformation is not invertible if we would like to
preserve these conditions unchanged on both manifolds. The only
condition for invertibility of pseudoduality is that pseudoduality
is between riemann normal coordinates with vanishing torsions. In
the orthonormal coframe method we have seen that our results are
similar to ones found before \cite{alvarez2, msarisaman2}. When we
consider the sigma models based on the Lie groups, we have seen that
pseudoduality in components imposes three different conditions; flat
space pseudoduality which yield that both $\lambda_{+}$ and $T$
vanish, and structures constants are the same, (anti)chiral
pseudoduality which yields ($\psi_{-} = 0$) $\psi_{+} = 0$ with
distinct structure constants. These conditions are the result of
equation (\ref{equa4.32}). For each case we found the conserved
super currents. We are going to use these results to find out the
pseudoduality conditions and conserved currents when applied to
symmetric spaces \cite{msarisaman3}.

\section*{Acknowledgments}

I would like to thank O. Alvarez for his comments, helpful
discussions, and reading an earlier draft of the manuscript. I would
like to thank E. A. Ivanov for bringing his important paper to my
attention.

\bibliographystyle{amsplain}

\end{document}